\documentclass[11pt]{article}
\usepackage[T1]{fontenc}
\usepackage[utf8]{inputenc}
\usepackage{lmodern}
\usepackage[a4paper,margin=22mm]{geometry}
\usepackage{amsmath,amssymb,graphicx,booktabs,array,longtable,tabularx}
\usepackage{setspace,placeins}
\usepackage[super,sort&compress]{natbib}
\setcitestyle{super,comma}
\usepackage[hidelinks]{hyperref}
\hypersetup{pdftitle={Coupling cellular trajectories and RNA kinetics on continuous expression manifolds with ManifoldVelo},pdfauthor={Zhen Zhou, Jiachen Li, Yuan Liu, Xiaoyong Pan, and Hong-Bin Shen},pdfsubject={Topology-conditioned RNA manifolds, detailed architecture and experimental results}}
\usepackage[font=small,labelfont=bf]{caption}
\raggedright
\newcommand{\R}{\mathbb{R}}
\newcommand{\spplus}{\operatorname{Softplus}}
\newcommand{\relu}{\operatorname{ReLU}}
\newcommand{\cmean}{\operatorname{mean}}
\newcommand{\dd}{\mathrm{d}}
\newcommand{\wrap}{\operatorname{wrap}}
\newcommand{\activeset}{\mathcal A}
\title{Coupling cellular trajectories and RNA kinetics on continuous expression manifolds with ManifoldVelo}
\author{Zhen Zhou, Jiachen Li, Yuan Liu, Xiaoyong Pan, and Hong-Bin Shen}
\date{29 October 2026}
\begin{document}
{\LARGE\bfseries Coupling cellular trajectories and RNA kinetics on continuous expression manifolds with ManifoldVelo\par}
\vspace{0.7em}
\begingroup
\setstretch{1.0}
\setlength{\parskip}{0.35em}
Zhen Zhou\textsuperscript{1}, Jiachen Li\textsuperscript{2}, Yuan Liu\textsuperscript{1}, Xiaoyong Pan\textsuperscript{1}, and Hong-Bin Shen\textsuperscript{1,*}\par
\small
\textsuperscript{1}Institute of Image Processing and Pattern Recognition, Shanghai Jiao Tong University, and Key Laboratory of System Control and Information Processing, Ministry of Education of China, Shanghai 200240, China.\par
\textsuperscript{2}Institute of Process Engineering, Chinese Academy of Sciences, Beijing 100190, China.\par
\textsuperscript{*}Correspondence: Hong-Bin Shen (\href{mailto:hbshen@sjtu.edu.cn}{hbshen@sjtu.edu.cn}).\par
\endgroup
\vspace{0.3em}

\begingroup
\setstretch{1.0}
\begin{abstract}
Jointly recovering pseudotime and RNA velocity from static single-cell transcriptomes requires reconciling global trajectory structure with local expression changes, particularly in complex cycles and branching differentiation. Here we introduce ManifoldVelo, a topology-conditioned framework that jointly learns cellular coordinates, continuous RNA trajectories and regulatory kinetics. Derivatives of unspliced and spliced RNA trajectories are matched to transcription--splicing--degradation dynamics, including between observed cells, allowing expression reconstruction, cellular ordering and kinetic estimation to inform one another. ManifoldVelo recovers pseudotime across all evaluated single-cell trajectory types, including cyclic, multibranch and smooth non-branching trajectories. ManifoldVelo achieves the best pseudotime recovery among compared methods in multipeak cyclic simulations and the highest directional agreement for spliced RNA velocity in both multipeak cyclic and branching simulations. Real-data analyses support phase prediction across experimental plates and differentiation trajectories consistent with pancreatic developmental stages and marker expression. Its central advantage is the joint recovery of temporal position and local dynamics across distinct trajectory structures within a reusable dynamic reference. New cells are assigned pseudotime and velocity predictions from spliced RNA alone, without refitting the reference model. ManifoldVelo provides a foundation for dynamic cell atlases that transfer learned temporal and kinetic information to new samples.
\end{abstract}
\endgroup
\vspace{-1em}

\section{Introduction}
Single-cell RNA sequencing captures heterogeneous populations as molecular snapshots, whereas the biological processes that generate this heterogeneity unfold continuously. Reconstructing these processes requires both the position of a cell along a trajectory and the direction in which its expression state is changing. Locally linear embedding (LLE) provides a foundational example of recovering low-dimensional manifold coordinates by preserving the local linear reconstruction relationships among high-dimensional observations\cite{lle2000}. Pseudotime methods established that cellular progression can be recovered from expression geometry: Monocle orders cells along reconstructed trajectories\cite{monocle2014}, and diffusion pseudotime uses diffusion distances to resolve progression and branching\cite{dpt2016}. Monocle 2 and Slingshot extend trajectory reconstruction to multiple lineages through graph learning and simultaneous principal curves, respectively\cite{monocle2017,slingshot2018}. PAGA represents connectivity between cellular populations\cite{paga2019}, while Palantir describes differentiation through probabilistic fate assignments\cite{palantir2019}. Comparative benchmarking has shown that the suitability of a trajectory method depends on the dataset and underlying topology\cite{saelens2019}, motivating representations that account explicitly for the structure of the biological process.

RNA velocity adds directional information by modeling the relationship between unspliced and spliced transcripts through transcription, splicing and degradation\cite{lamanno2018}. scVelo generalized this framework to transient populations by fitting dynamical splicing kinetics\cite{scvelo2020}. Subsequent methods have expanded the treatment of kinetic heterogeneity and uncertainty. cellDancer and DeepVelo use neural models to estimate cell-dependent kinetics\cite{celldancer2024,deepvelo2024}; veloVI uses a deep generative model to infer transcriptional dynamics and quantify uncertainty\cite{velovi2024}; and cell2fate organizes transcriptional changes into dynamic modules\cite{cell2fate2025}. These approaches show how richer kinetic models can connect local RNA changes with the heterogeneous states observed during a biological transition.

Several methods also integrate temporal coordinates with expression dynamics. UniTVelo models spliced-expression profiles along a unified cellular time and derives the corresponding unspliced dynamics\cite{unitvelo2022}. LatentVelo couples a variational representation of RNA expression to structured neural ordinary differential equations, using a latent regulatory state to represent lineage-dependent dynamics\cite{latentvelo2023}. scTour jointly estimates pseudotime, a latent vector field and a transcriptomic representation, and predicts dynamics for additional cells or datasets\cite{sctour2023}. At the level of fate inference, CellRank combines directional information with transcriptomic similarity to estimate fate probabilities\cite{cellrank2022}. Together, these studies establish that cellular time, representation and dynamics can inform one another; the choice of how to couple them determines which aspects of a trajectory are constrained by the model.

Cyclic processes make this coupling particularly explicit because progression must return to the same expression state. DeepCycle learns continuous cell-cycle progression from spliced and unspliced expression patterns\cite{deepcycle2022}, while tricycle transfers cell-cycle position from a fixed reference embedding\cite{tricycle2022}. VeloCycle introduced a probabilistic manifold-constrained RNA velocity framework that links expression, kinetic parameters and progression speed; its formulation also includes extensions beyond periodic trajectories\cite{velocycle2024}. GraphVelo uses local tangent-space constraints to refine supplied velocities and transport them between representations\cite{graphvelo2025}. These complementary formulations motivate a closer connection between an explicit continuous RNA trajectory and the kinetic field that describes motion along it, especially when expression has several peaks within a cycle or when progenitors contribute to multiple lineages.

Regulatory and multimodal models provide further biological structure. TFvelo estimates RNA dynamics from transcription-factor regulation without requiring a spliced/unspliced decomposition\cite{tfvelo2024}. TSvelo and RegVelo couple regulatory interactions to transcription and splicing\cite{tsvelo2026,regvelo2026}. MultiVelo links chromatin accessibility with RNA production and processing\cite{multivelo2023}, and MultiVeloVAE extends dynamical inference across lineages, samples and partially overlapping modalities\cite{multivelovae2025}. ArchVelo represents transcriptional inputs using chromatin-derived archetypal programs\cite{archvelo2026}. Within an RNA-based analysis, these developments motivate a specific modeling question: how can cellular coordinates and explicit trajectories of both RNA species be refined together under a regulatory kinetic field, while encoding periodic closure and a shared origin for branching lineages?

Here we introduce ManifoldVelo, a framework that couples cellular trajectories with RNA kinetics on continuous expression manifolds. Each cell is assigned a coordinate on a differentiable curve that reconstructs its unspliced and spliced expression. Collocation residuals constrain the derivatives of both reconstructed RNA species to agree with a transcription--splicing--degradation field, including between observed cells. RNA reconstruction and kinetic consistency therefore jointly inform the fitted trajectory and cellular coordinates. The temporal domain supplies the structural constraints: periodic decoders close cyclic trajectories, whereas overlapping lineage curves share a decoded root state and retain distinct terminal paths. A single open curve represents an unbranched process. These representations place cyclic and branching dynamics within a common fitting principle, with architecture choices matched to the configured topology.

The fitted expression manifold also provides a dynamic reference for additional cells measured with spliced RNA alone. Reference transfer and prediction for new cells have been developed for cell-cycle position and latent dynamics\cite{tricycle2022,sctour2023}. Here, a new S-only cell is mapped onto the fixed U/S reference, after which the model returns its temporal coordinate, reconstructed RNA state and gene-space kinetic velocity. The reference parameters remain fixed during this step. We evaluate this capability with held-out simulation cells and a plate-held-out U2OS protocol, linking the learned continuous trajectory to prediction beyond the cells used for paired-U/S fitting.

Across multipeak cyclic simulations, ManifoldVelo achieved circular time errors of 0.64\% and 7.40\%, together with held-out S-velocity cosines of 0.9776 and 0.9421. Shared-root lineage models recovered velocity cosines of 0.9735 and 0.9404 in trifurcation and nested-branch simulations. In the frozen real-data comparison, phase error decreased by 47.0\% in WT2 and 8.7\% in plate-held-out U2OS relative to the recorded VeloCycle results. Matched component ablations further supported the contributions of U dynamics, expression neighborhoods and joint fitting to phase recovery and local velocity coherence. Together, these experiments establish the utility of jointly fitting cellular position and RNA dynamics on explicit, topology-conditioned expression trajectories.

\section{Results}
\subsection{A continuous RNA manifold couples cellular position and local dynamics}
\begin{figure}[!p]
\centering
\includegraphics[width=\textwidth,height=.83\textheight,keepaspectratio]{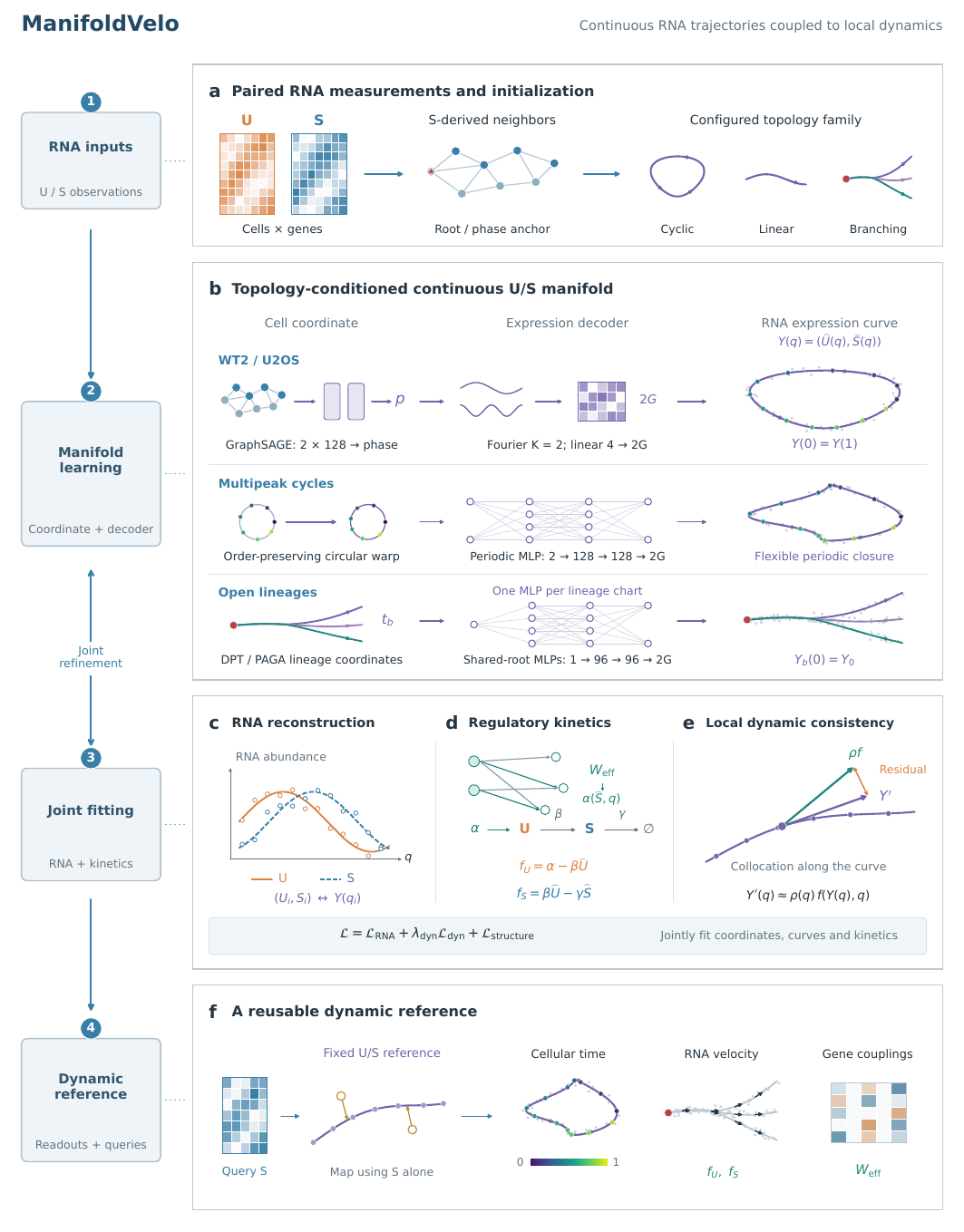}
\caption{\textbf{ManifoldVelo jointly reconstructs cellular trajectories and RNA dynamics.}
\textbf{a}, Paired unspliced (U) and spliced (S) matrices, S-derived neighborhoods and an anchor initialize the configured topology.
\textbf{b}, Three coordinate--decoder configurations learn periodic or shared-root expression curves; widths include input, hidden and output dimensions.
\textbf{c}, Both RNA species constrain reconstruction.
\textbf{d}, Masked gene couplings modulate transcription in the U/S kinetic field.
\textbf{e}, Curve derivatives and kinetic vectors are matched at collocation points, jointly refining coordinates, curves and kinetics.
\textbf{f}, The fitted reference provides cellular time, RNA velocity and gene couplings; S-only query cells map to it without refitting. $G$ denotes the number of modeled genes. All panels show conceptual schematics.}
\label{fig:framework}
\end{figure}

ManifoldVelo jointly fits a continuous expression trajectory and a local RNA kinetic field (Fig.~\ref{fig:framework}). At temporal coordinate $q$, the decoder returns both RNA species, $Y(q)=(\widehat U(q),\widehat S(q))$. Matching the curve tangent to the kinetic field connects cellular position to expression-change direction. Jointly updating coordinates, curves and kinetics allows RNA dynamics to refine temporal ordering and constrains the field between observed cells.

The temporal domain specifies trajectory structure (Fig.~\ref{fig:framework}b). A circular domain with a periodic decoder enforces closure. Open root-to-terminal paths define overlapping lineage charts whose decoders share an exact root state; shared-cell constraints link their progenitor regions. One chart gives the unbranched case. The benchmark presets specify the cyclic or open family, and expression-derived paths determine open lineages. These constraints act on RNA expression functions; UMAP displays the inferred positions and velocities.

WT2 and U2OS use two 128-dimensional GraphSAGE layers, a two-dimensional circular head and a second-order Fourier decoder. Multipeak cycles use an order-preserving circular map and a periodic neural decoder. Open trajectories use lineage-specific decoders with two hidden layers of width 96 and a shared root parameter. Table~\ref{tab:architecture} gives all dimensions. The regulatory transcription--splicing--degradation component builds on TSvelo\cite{tsvelo2026}; explicit RNA curves and tangent--kinetic consistency connect it to the temporal manifold.

The fitted model supplies a reusable reference for S-only queries through graph encoding, circular mapping or open-curve projection. Each route predicts time, reconstructed U/S and velocity without refitting. Native kinetic vectors and decoder tangents are exported separately, preserving the distinction between the learned field and the curve derivative.

\FloatBarrier

\subsection{Automatic anchoring supports recovery across simulated trajectory geometries}
\begin{figure}[p]
\centering
\includegraphics[width=\textwidth,height=.65\textheight,keepaspectratio]{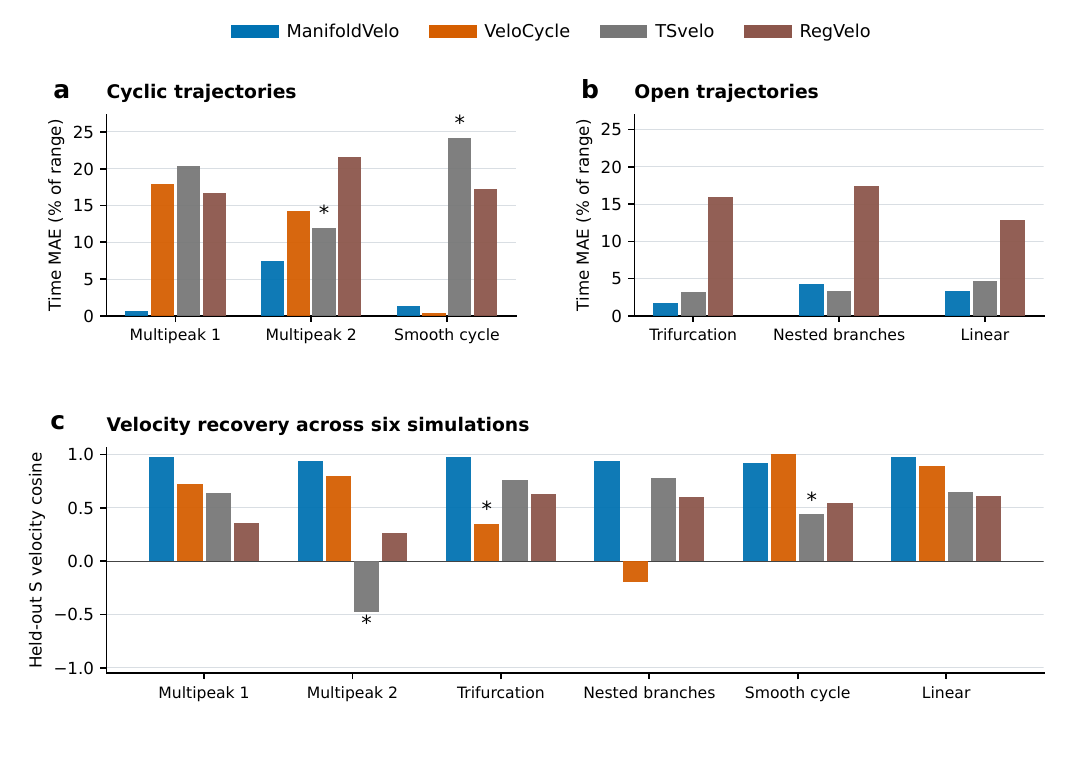}
\caption{\textbf{Time and velocity recovery across six simulations with automatically inferred ManifoldVelo anchors.}
\textbf{a}, Circular time error for multipeak data1, data2 and a smooth cycle. \textbf{b}, Linear time error for trifurcation, nested branches and a single trajectory. VeloCycle produces a periodic phase in these open stress tests, so a linear time error is not defined for its native output. \textbf{c}, Held-out S-velocity cosine in all six systems, including those VeloCycle stress tests. Values are single archived runs with training seed 0; there are no seed-based error bars. ManifoldVelo uses training U/S for automatic selection and S-only query input. The displayed VeloCycle configurations use six harmonics plus ridge adaptation for the multipeak cases and two harmonics for the remaining cases. TSvelo and RegVelo receive no supplied regulatory-edge network in these simulations. Asterisks retain the recorded budget/stopping qualifications, including the unconverged data2 TSvelo checkpoint. All archived methods, fitting-status flags, S-only readout adaptations and metric explanations are retained in the accompanying source-data archive. Numerical source data are provided with the figure.}
\label{fig:simulation}
\end{figure}

We tested six simulated systems spanning two multipeak cycles, a smooth cycle, a linear trajectory, a trifurcation and nested branches. Each experiment supplied paired U/S training cells and S-only query cells. The complete ManifoldVelo pipeline inferred a root candidate from training U/S, rebuilt lineage charts where applicable, fitted the fixed model configuration and exported both time and velocity. Reference time and query U were reserved for evaluation. Figure~\ref{fig:simulation} summarizes the four principal model comparisons; Table~\ref{tab:simulation} compares recovery with the strongest recorded alternative for each endpoint. Supplementary Table S7 provides the complete method comparison, resource settings and readout qualifications. Supplementary Tables S12 and S13 document the time-error and velocity-difference calculations used in the comparisons.

We next examined the two multipeak cycles and the two branching systems in detail (Figs.~\ref{fig:simulation_cycles} and \ref{fig:simulation_branches}). These experiments test complementary properties of the representation: continuity around a closed temporal domain and progression from a shared root into distinct lineages. The linear control gave a time error of 3.32\% and a held-out S-velocity cosine of 0.9757 for ManifoldVelo. The smooth cyclic control favored VeloCycle, with a time error of 0.41\% and velocity cosine of 0.9994, compared with 1.35\% and 0.9223 for ManifoldVelo. These controls place the complex-trajectory results alongside simpler geometries.

Across the six simulations, the complete ManifoldVelo recovery took 23.4--91.3\,s per dataset under the recorded one-GPU, one-worker configuration. Every query cell received a finite time estimate and velocity. The original UMAP coordinates were retained across methods, with query cells outlined and automatically selected anchors marked. The geometry of the display therefore remained fixed while model outputs changed.

\begin{table}[!htbp]
\centering\small
\caption{\textbf{Recovery across six simulations and the strongest recorded comparators.}
Time error and S-velocity cosine are evaluated on S-only queries. The comparator is chosen independently for each endpoint; bold indicates the better score.}
\label{tab:simulation}
\setlength{\tabcolsep}{3pt}
\begin{tabularx}{\textwidth}{@{}l r X r X r@{}}
\toprule
 & \multicolumn{2}{c}{Time MAE (\%)} & \multicolumn{2}{c}{S-velocity cosine} & Ours \\
\cmidrule(lr){2-3}\cmidrule(lr){4-5}
Dataset & Ours & Best comparator & Ours & Best comparator & Seconds \\
\midrule
Multipeak data1 & \textbf{0.64} & 11.92 (cellDancer$^{\ddagger}$) & \textbf{0.9776} & 0.8208 (scVelo (stoch.)) & 83.3 \\
Multipeak data2 & \textbf{7.40} & 11.93 (TSvelo$^{*}$) & \textbf{0.9421} & 0.8173 (scVelo (stoch.)) & 91.3 \\
Smooth cycle & 1.35 & \textbf{0.41} (VeloCycle) & 0.9223 & \textbf{0.9994} (VeloCycle) & 23.4 \\
Trifurcation & \textbf{1.75} & 3.17 (TSvelo) & \textbf{0.9735} & 0.7624 (TSvelo) & 74.9 \\
Nested branches & 4.28 & \textbf{3.33} (TSvelo) & \textbf{0.9404} & 0.7745 (TSvelo) & 86.1 \\
Linear & \textbf{3.32} & 4.65 (TSvelo) & \textbf{0.9757} & 0.8917 (VeloCycle) & 31.8 \\
\bottomrule
\end{tabularx}
\par\smallskip\footnotesize
All runs use seed 0. ManifoldVelo infers its anchor from training U/S. $^{*}$The data2 TSvelo result is the archived unconverged checkpoint. $^{\ddagger}$Data1 cellDancer uses the common velocity-pseudotime adapter after its native time readout failed. TFvelo's total-RNA velocity is scored separately. The periodic VeloCycle open-trajectory stress tests contribute velocity scores, not linear-time errors. All method results, qualifications and resource settings appear in the Supplementary Information. ManifoldVelo recovery times include automatic anchoring, fitting and output, using one numerical CPU thread and one RTX 4090 D GPU per job.
\end{table}

\FloatBarrier

\subsection{Flexible periodic manifolds recover phase and velocity in multipeak cycles}
\begin{figure}[p]
\centering
\includegraphics[width=\textwidth,height=.70\textheight,keepaspectratio]{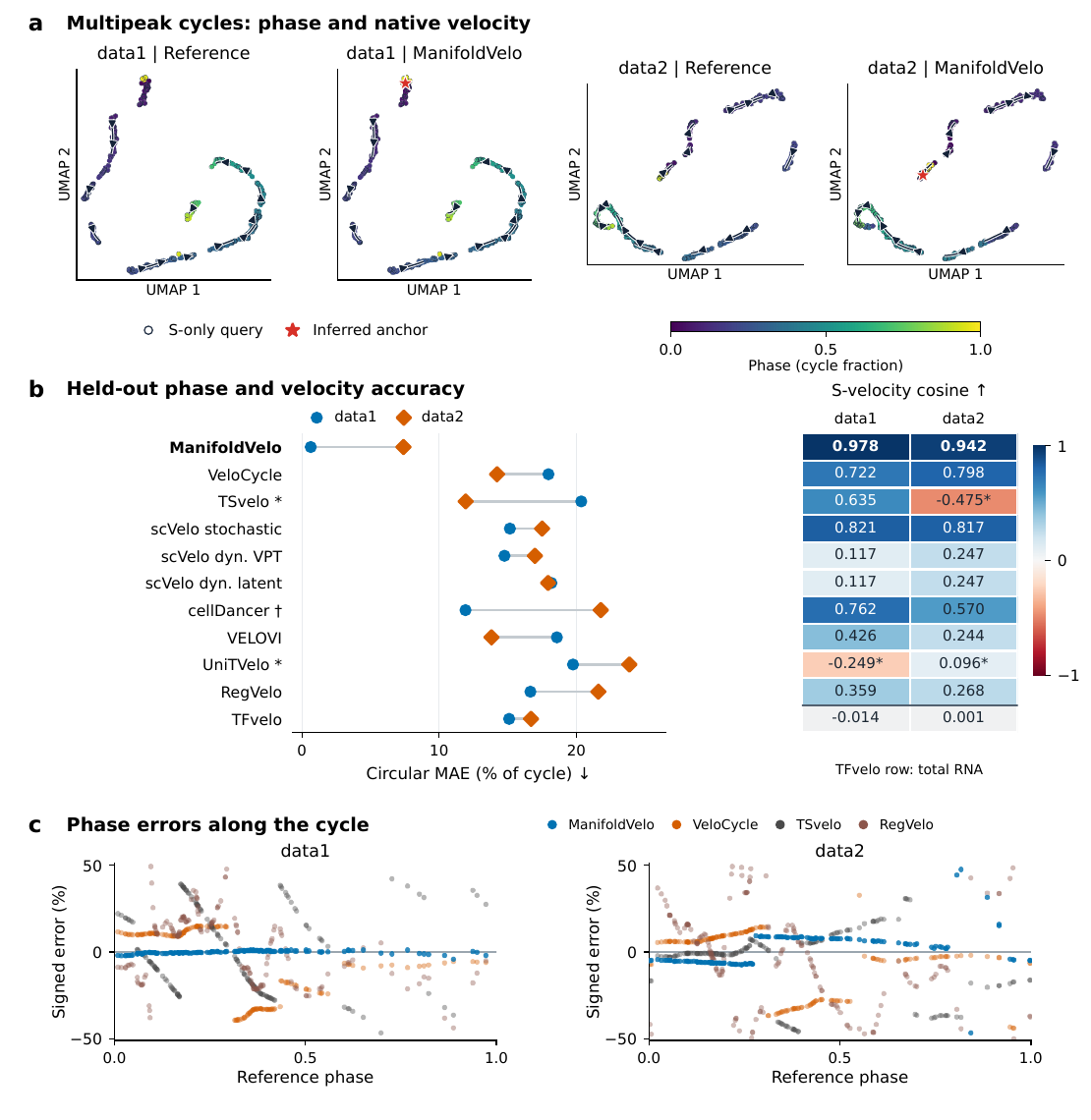}
\caption{\textbf{Phase and velocity recovery in two multipeak cyclic simulations.}
\textbf{a}, Reference phase and true S velocity, and ManifoldVelo phase and native S velocity, on each dataset's original UMAP. All 480 cells are colored; thin outlines identify the 160 S-only queries and red stars identify automatically inferred training anchors. \textbf{b}, Query circular MAE (paired dots) and mean S-velocity cosine (heatmap) for eleven method readouts. The separately boxed TFvelo row reports total-RNA cosine against the total-RNA reference. \textbf{c}, Signed circular phase errors for individual queries, after the archived training-only phase calibration. Velocity directions are preserved independently of color calibration. Values use training seed 0. Asterisks mark the unconverged data2 TSvelo checkpoint and default-budget UniTVelo runs; the dagger marks the common velocity-pseudotime readout used for data1 cellDancer after its native pseudotime procedure failed. scVelo stochastic uses $k=30$; VPT denotes velocity pseudotime. Quantitative velocities are evaluated in gene space before display projection or smoothing. Numerical and per-cell source data accompany the figure.}
\label{fig:simulation_cycles}
\end{figure}

Multipeak data1 and data2 each comprised 320 paired U/S training cells and 160 S-only query cells across 18 genes. ManifoldVelo inferred the phase anchor from training RNA and fitted a periodic neural expression curve. Figure~\ref{fig:simulation_cycles} combines the reference and recovered fields on the fixed UMAP with held-out errors for all eleven method readouts. The phase-residual panels additionally resolve the errors along the reference cycle.

For data1, ManifoldVelo achieved a circular time error of 0.64\%, compared with 17.97\% for VeloCycle and 20.36\% for TSvelo. For data2, the corresponding errors were 7.40\%, 14.22\% and 11.93\%. Relative to the recorded VeloCycle configurations, these represent reductions of 96.4\% and 48.0\% in circular error. ManifoldVelo had the lowest held-out time error among the eleven displayed readouts in both datasets.

The phase improvements were accompanied by more accurate high-dimensional RNA velocities. The held-out S-velocity cosine was 0.9776 in data1 and 0.9421 in data2, compared with 0.7224 and 0.7985 for VeloCycle. The strongest remaining comparator for this metric was scVelo stochastic, with 0.8208 and 0.8173. TSvelo gave 0.6346 and $-0.4747$, respectively, with its data2 value obtained from the recorded unconverged checkpoint. Thus, within these multipeak simulations, the periodic manifold recovered both cellular position and the direction of expression change from S-only query observations.

\subsection{Shared-root lineage manifolds recover directional dynamics across branches}
\begin{figure}[p]
\centering
\includegraphics[width=\textwidth,height=.70\textheight,keepaspectratio]{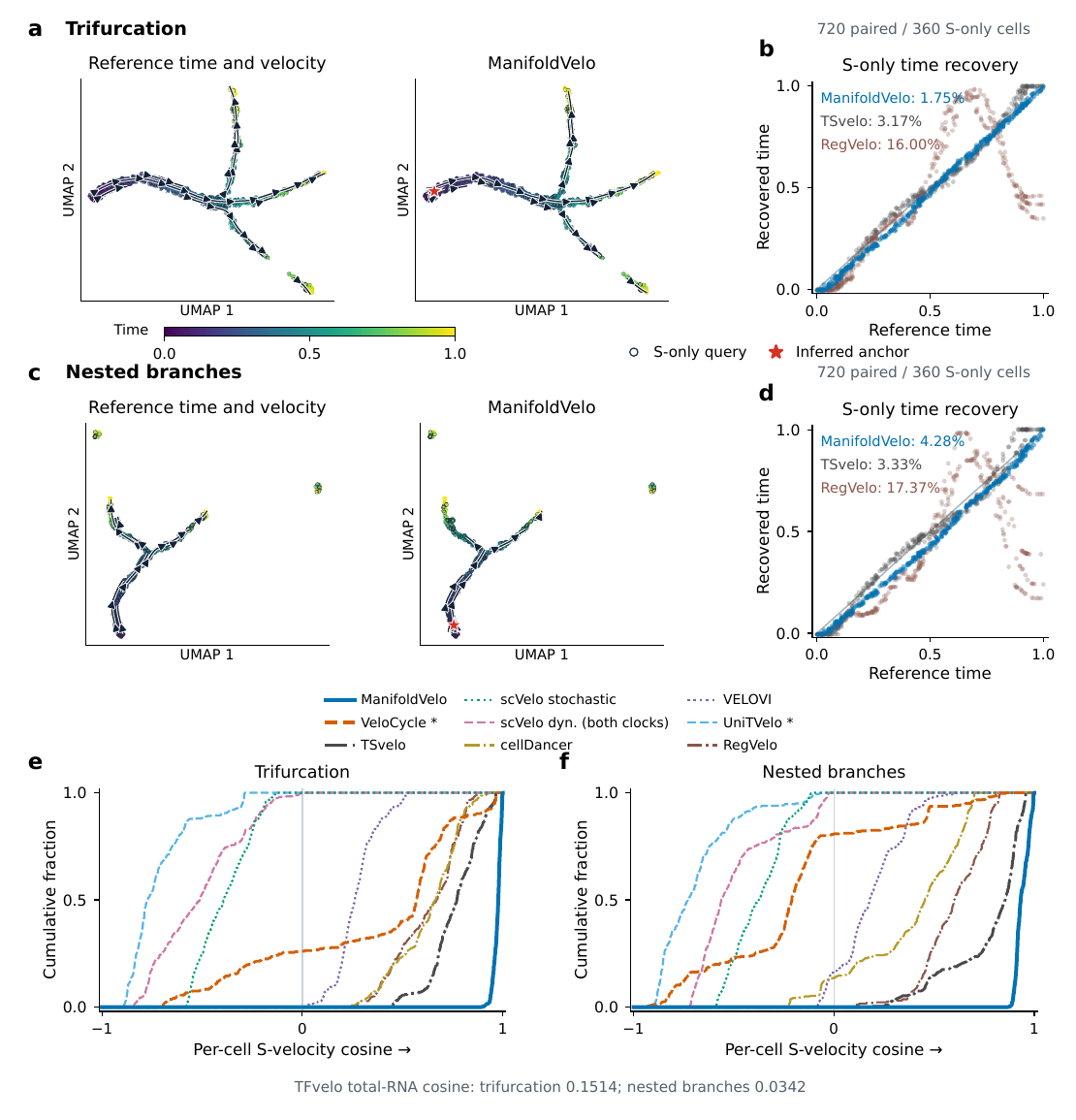}
\caption{\textbf{Temporal progression and cell-level velocity accuracy in branching simulations.}
\textbf{a,c}, Reference time and true S velocity alongside ManifoldVelo outputs for trifurcation and nested branches. All 1,080 cells are colored on the original UMAP; thin outlines mark 360 S-only queries and red stars mark automatically inferred training roots. \textbf{b,d}, Query time estimates against reference time, with a diagonal identity line and MAE annotations. \textbf{e,f}, Empirical cumulative distributions of query S-velocity cosine; distributions concentrated toward 1 indicate greater directional accuracy. The two scVelo dynamical clocks share one velocity field. TFvelo's total-RNA means are reported separately below the panels. VeloCycle retains its periodic output in these open-trajectory stress tests, so it contributes a velocity curve but no linear-time error. Asterisks identify the fixed-budget trifurcation VeloCycle fit and default-budget UniTVelo runs. ManifoldVelo uses its fixed frozen training budget. All runs use seed 0; velocity metrics use unsmoothed gene-space predictions. Numerical and per-cell source data accompany the figure.}
\label{fig:simulation_branches}
\end{figure}

We evaluated trifurcation and nested-branch simulations to test recovery beyond a periodic temporal domain. Each system contained 720 paired training cells and 360 S-only queries across 36 genes. The automatic-root pipeline constructed overlapping lineage charts and fitted expression curves with a shared decoded root. Figure~\ref{fig:simulation_branches} shows the reference and recovered fields together with cell-level time agreement and distributions of high-dimensional velocity accuracy.

In the trifurcation, ManifoldVelo achieved a time error of 1.75\% and an S-velocity cosine of 0.9735, compared with 3.17\% and 0.7624 for TSvelo. In nested branches, ManifoldVelo retained a velocity cosine of 0.9404, exceeding TSvelo's 0.7745, while TSvelo had the lower time error: 3.33\% compared with 4.28\%. Temporal position and local expression-change direction therefore provide complementary information about recovery on a branching trajectory.

The per-cell velocity distributions show that the high mean accuracy extended across the query populations. Cosines exceeded 0.9 in 359 of 360 trifurcation queries (99.7\%) and 331 of 360 nested-branch queries (91.9\%) for ManifoldVelo. Its mean S-velocity cosine was highest among the displayed S-velocity methods in both systems. Applying the recorded periodic VeloCycle implementation to these open trajectories gave stress-test cosines of 0.3485 and $-0.1922$, respectively. These runs evaluate that periodic configuration; VeloCycle also describes non-periodic extensions\cite{velocycle2024}. Together with the multipeak cyclic results, these experiments demonstrate accurate recovery of RNA dynamics across closed trajectories and multiple root-to-terminal paths using topology-specific expression manifolds.

\FloatBarrier

\subsection{WT2 and U2OS show improved periodic phase recovery}
We evaluated periodic phase recovery in the WT2 mouse neurosphere sample and human U2OS FUCCI cells. WT2 contained 7,057 cells with a 391-gene panel and RNA-derived tricycle phase annotations\cite{tricycle2022}. U2OS contained 1,152 cells across three plates, with 1,146 available FUCCI phase labels and a 107-gene ManifoldVelo panel\cite{mahdessian2021}. In the frozen comparison, U2OS was evaluated by leaving one plate out and querying it with S alone. The predefined roots and model configurations were retained throughout these comparisons.

ManifoldVelo achieved a circular mean absolute error of $3.762\pm0.065$\% of a cycle in WT2, compared with 7.102\% for VeloCycle. In U2OS, the three-plate S-only evaluation gave $11.481\pm0.075$\%, compared with 12.575\% for VeloCycle. These changes correspond to relative error reductions of 47.0\% and 8.7\%, respectively (Table~\ref{tab:frozen}). The standard deviations summarize three optimization seeds. WT1, the companion neurosphere replicate, is included in the same table and favored VeloCycle, identifying variation in the benefit across samples.

\begin{figure}[p]
\centering
\includegraphics[width=\textwidth,height=.80\textheight,keepaspectratio]{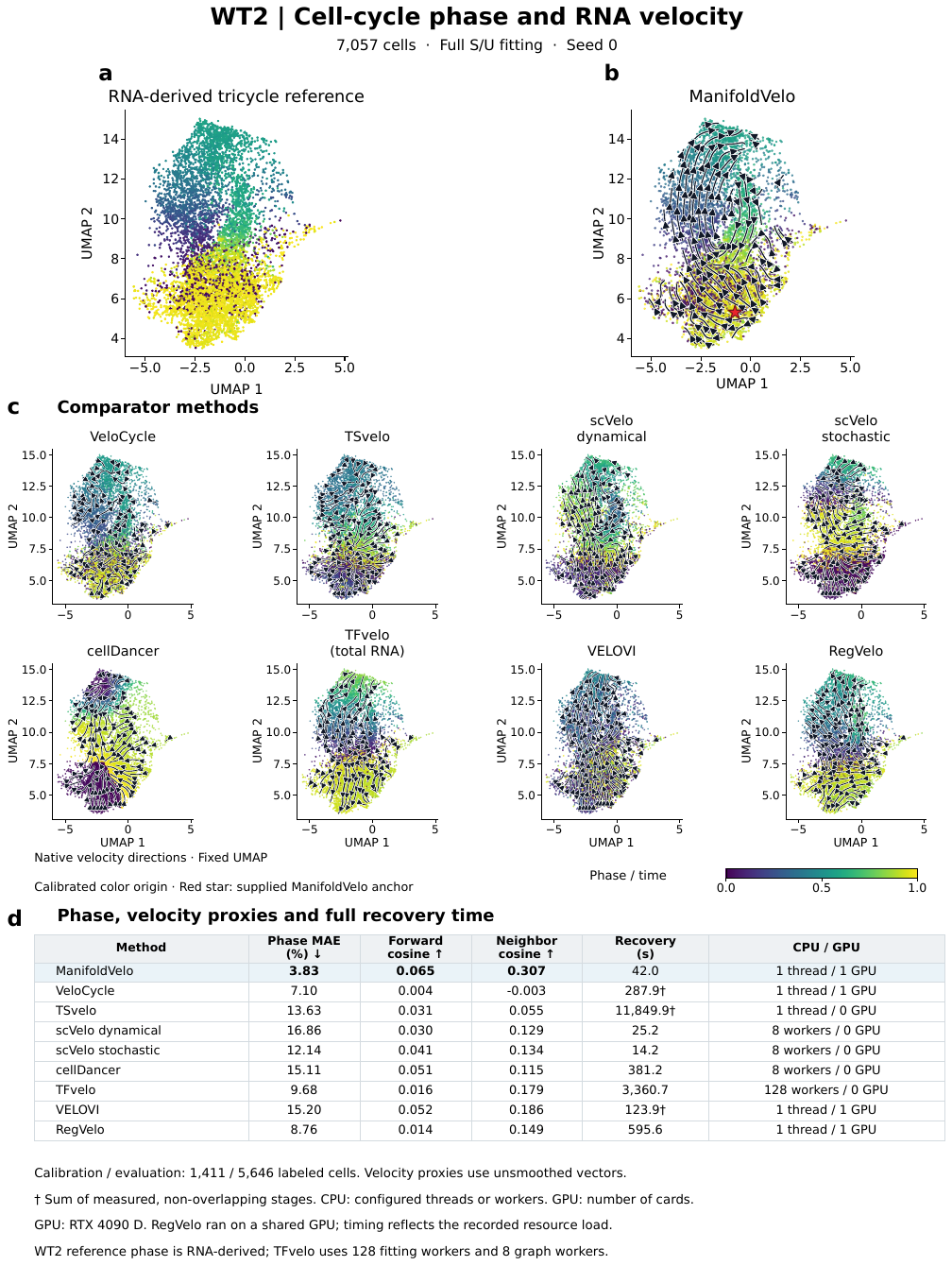}
\caption{\textbf{WT2 phase and RNA velocity under the full-S/U visualization protocol.}
\textbf{a}, RNA-derived tricycle reference. \textbf{b}, ManifoldVelo. \textbf{c}, Eight comparator modes. \textbf{d}, Phase error, velocity proxies and fitting-inclusive recovery time. All 7,057 cells are modeled with seed 0; 1,411 labels calibrate phase and 5,646 evaluate it. The original UMAP and native velocity directions are retained; phase colors receive a global shift. The star marks the supplied ManifoldVelo anchor. TFvelo displays total-RNA velocity; other panels show S velocity. Scores use unsmoothed vectors. Daggers identify summed runtime stages; parallelism is shown per row. This protocol is separate from Table~\ref{tab:frozen}.}
\label{fig:wt2}
\end{figure}

\begin{figure}[p]
\centering
\includegraphics[width=\textwidth,height=.80\textheight,keepaspectratio]{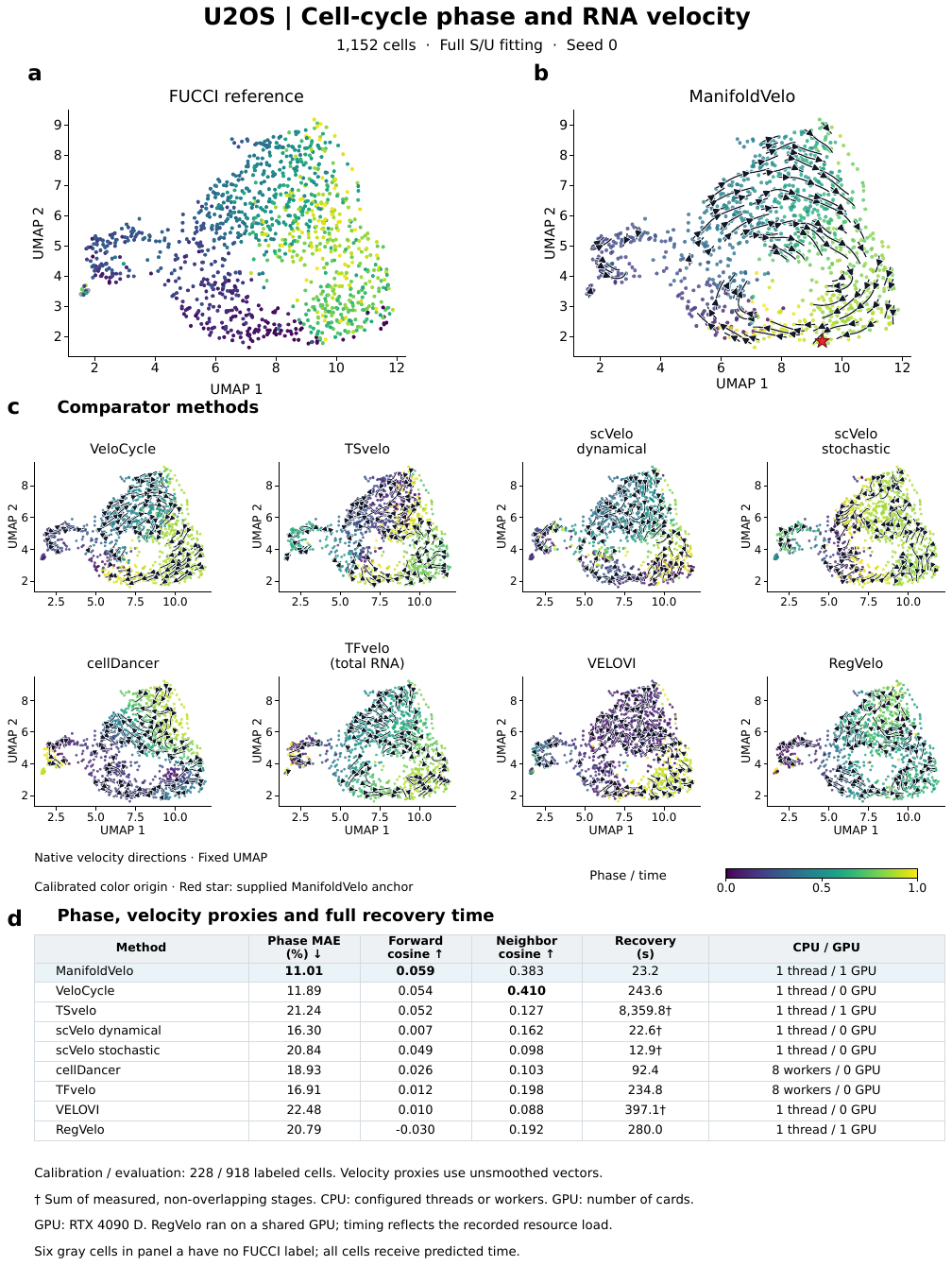}
\caption{\textbf{U2OS phase and RNA velocity under the full-S/U visualization protocol.}
\textbf{a}, FUCCI reference; six unlabeled cells remain gray. \textbf{b}, ManifoldVelo predictions for all 1,152 cells. \textbf{c}, Eight comparator modes. \textbf{d}, Quantitative scores and recovery times. All-cell S/U fitting uses seed 0, with 228 labels for phase calibration and 918 for evaluation. The original UMAP and native velocity directions are retained; phase colors receive a global shift. The star marks the supplied anchor. TFvelo displays total-RNA velocity. Scores use unsmoothed vectors; daggers identify summed runtime stages. The plate-held-out S-only evaluation is reported separately in Table~\ref{tab:frozen}.}
\label{fig:u2os}
\end{figure}

To display phase and velocity from the same fitted cells, we also prepared a separately defined full-S/U comparison. Its phase calibration used a global rotation without reference-driven reversal, and all panels used the original UMAP embedding (Figs.~\ref{fig:wt2} and \ref{fig:u2os}). In this protocol, phase errors were 3.834\% for WT2 and 11.012\% for U2OS. Forward velocity cosines in the display space were 0.0646 and 0.0590, compared with 0.0037 and 0.0539 for VeloCycle. Neighbor velocity coherence was higher for ManifoldVelo in WT2 and higher for VeloCycle in U2OS. These panels connect the phase advantage to the recovered local flow while retaining the distinct information provided by direction and smoothness.

The full-S/U panels include scVelo dynamical and stochastic modes, cellDancer, TFvelo and veloVI, together with TSvelo, VeloCycle and RegVelo. Each method colors all cells with its native temporal output after the stated calibration. In U2OS, the six cells lacking FUCCI labels remain predicted and are uncolored only in the reference panel. Tables~\ref{tab:full_WT2} and \ref{tab:full_U2OS} report the numerical values, fitting-inclusive recovery times and computational parallelism for this visualization comparison.

\begin{table}[!htbp]
\centering
\caption{\textbf{Periodic phase recovery in the frozen real-data comparison.}
Circular mean absolute error is reported as a percentage of one cycle (lower is better). ManifoldVelo values are means $\pm$ sample standard deviations across seeds 0, 1 and 2; comparator values are single archived runs. Bold values identify the lowest error in each row.}
\label{tab:frozen}
\small
\setlength{\tabcolsep}{5pt}
\begin{tabular}{lrrrr}
\toprule
Dataset & Evaluated cells & ManifoldVelo & VeloCycle & TSvelo \\
\midrule
WT1 & 4599 & 5.301 $\pm$ 0.094 & \textbf{3.337} & 10.921 \\
WT2 & 5646 & \textbf{3.762 $\pm$ 0.065} & 7.102 & 13.626 \\
U2OS & 1146 & \textbf{11.481 $\pm$ 0.075} & 12.575 & 15.408 \\
\bottomrule
\end{tabular}
\par\vspace{1em}
\begin{minipage}{\textwidth}\footnotesize
WT1 and WT2 use RNA-derived tricycle references. U2OS uses three plate-held-out S-only folds with FUCCI labels. ManifoldVelo retains the recorded roots; comparators retain their original protocols. Calibration uses the archived rotation/reflection rule. Method-specific input handling is documented in Methods. This table is separate from the full-S/U velocity visualizations.
\end{minipage}
\end{table}

\begin{table}[!htbp]
\centering\small
\caption{\textbf{WT2 full-S/U phase, velocity and recovery time.}
Phase error is a percentage of one cycle; forward and neighbor scores are fixed-UMAP velocity proxies. Each row uses the recorded seed-0 prediction.}
\label{tab:full_WT2}
\setlength{\tabcolsep}{3.5pt}
\begin{tabular}{@{}lrrrrlr@{}}
\toprule
Method & Phase (\%) & Forward & Neighbor & Seconds & CPU & GPUs \\
\midrule
ManifoldVelo & 3.834 & 0.0646 & 0.3069 & 42.0 & 1 thread & 1 \\
VeloCycle & 7.102 & 0.0037 & -0.0035 & 287.9$^{\dagger}$ & 1 thread & 1 \\
TSvelo & 13.626 & 0.0314 & 0.0552 & 11849.9$^{\dagger}$ & 1 thread & 0 \\
scVelo dynamical & 16.865 & 0.0299 & 0.1286 & 25.2 & 8 workers & 0 \\
scVelo stochastic & 12.143 & 0.0409 & 0.1339 & 14.2 & 8 workers & 0 \\
cellDancer & 15.107 & 0.0509 & 0.1150 & 381.2 & 8 workers & 0 \\
TFvelo & 9.679 & 0.0163 & 0.1790 & 3360.7 & 128 workers & 0 \\
VELOVI & 15.195 & 0.0519 & 0.1862 & 123.9$^{\dagger}$ & 1 thread & 1 \\
RegVelo & 8.764 & 0.0136 & 0.1495 & 595.6 & 1 thread & 1 \\
\bottomrule
\end{tabular}
\par\smallskip\footnotesize
$\dagger$: sum of measured, non-overlapping fitting and output stages; CPU parallelism and GPU count are shown in the same row. All timings include fitting from prepared RNA input. RegVelo timings were recorded on a shared GPU. TFvelo reports total-RNA velocity. Methods retain their recorded gene panels and preprocessing. Display smoothing is excluded from scores.
\end{table}

\begin{table}[!htbp]
\centering\small
\caption{\textbf{U2OS full-S/U phase, velocity and recovery time.}
Phase error is a percentage of one cycle; forward and neighbor scores are fixed-UMAP velocity proxies. Each row uses the recorded seed-0 prediction.}
\label{tab:full_U2OS}
\setlength{\tabcolsep}{3.5pt}
\begin{tabular}{@{}lrrrrlr@{}}
\toprule
Method & Phase (\%) & Forward & Neighbor & Seconds & CPU & GPUs \\
\midrule
ManifoldVelo & 11.012 & 0.0590 & 0.3829 & 23.2 & 1 thread & 1 \\
VeloCycle & 11.891 & 0.0539 & 0.4104 & 243.6 & 1 thread & 0 \\
TSvelo & 21.237 & 0.0520 & 0.1270 & 8359.8$^{\dagger}$ & 1 thread & 1 \\
scVelo dynamical & 16.300 & 0.0067 & 0.1625 & 22.6$^{\dagger}$ & 1 thread & 0 \\
scVelo stochastic & 20.839 & 0.0493 & 0.0984 & 12.9$^{\dagger}$ & 1 thread & 0 \\
cellDancer & 18.933 & 0.0264 & 0.1026 & 92.4 & 8 workers & 0 \\
TFvelo & 16.910 & 0.0117 & 0.1984 & 234.8 & 8 workers & 0 \\
VELOVI & 22.477 & 0.0096 & 0.0880 & 397.1$^{\dagger}$ & 1 thread & 0 \\
RegVelo & 20.789 & -0.0301 & 0.1919 & 280.0 & 1 thread & 1 \\
\bottomrule
\end{tabular}
\par\smallskip\footnotesize
$\dagger$: sum of measured, non-overlapping fitting and output stages; CPU parallelism and GPU count are shown in the same row. All timings include fitting from prepared RNA input. RegVelo timings were recorded on a shared GPU. TFvelo reports total-RNA velocity. Methods retain their recorded gene panels and preprocessing. Display smoothing is excluded from scores.
\end{table}

\FloatBarrier

\subsection{Dynamics, neighborhood information and joint fitting improve phase recovery and local coherence}
\begin{figure}[p]
\centering
\includegraphics[width=\textwidth,height=.67\textheight,keepaspectratio]{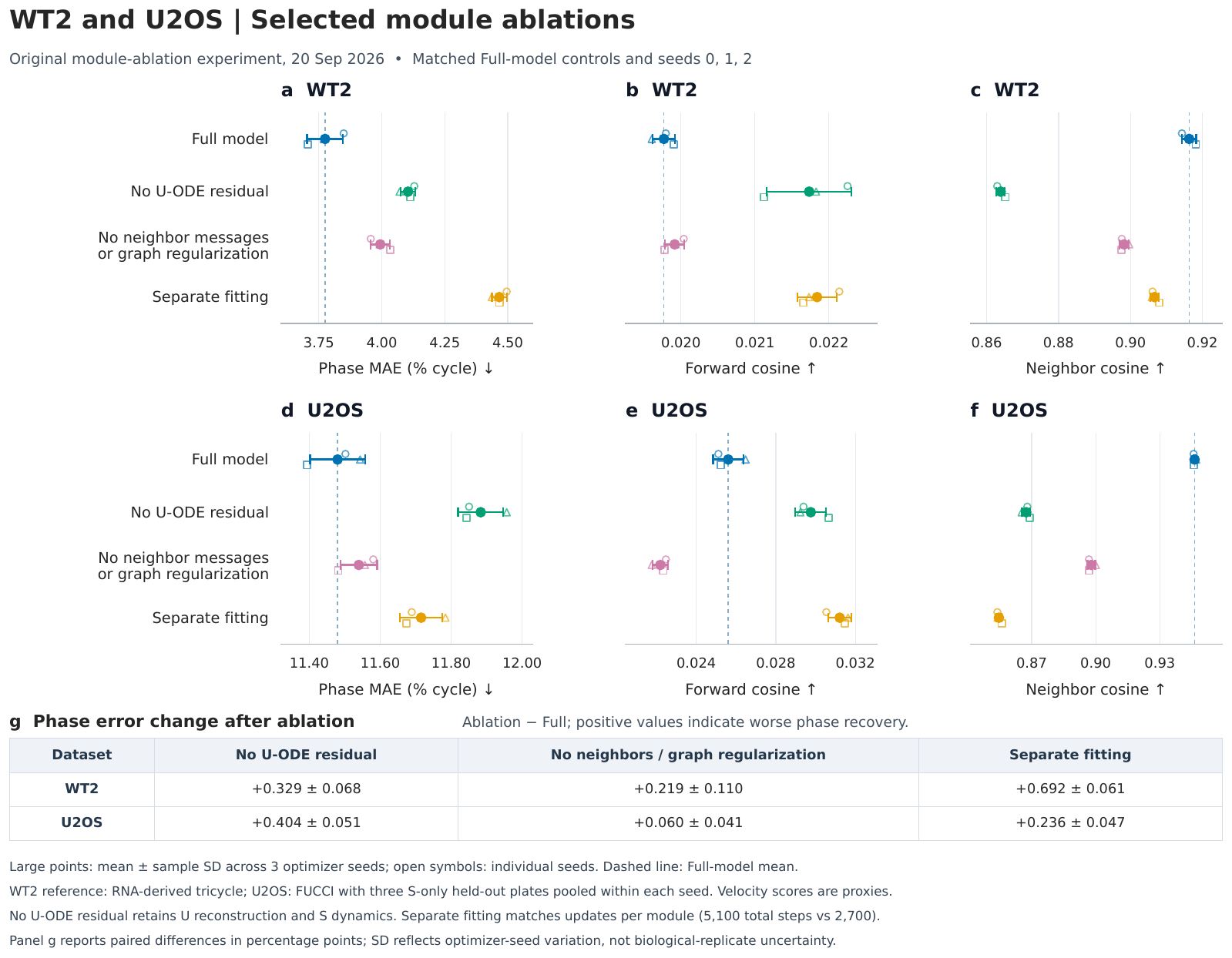}
\caption{\textbf{Component ablations link dynamics, neighborhoods and joint fitting to phase recovery and velocity coherence.}
\textbf{a--f}, Phase error, forward-reference cosine and neighbor velocity cosine for WT2 and U2OS. Large points and error bars show means and sample standard deviations across three optimization seeds; open symbols show every seed. Dashed lines indicate the full-model mean. \textbf{g}, Paired phase-error changes relative to the full model. The displayed controls remove only the U-ODE residual, jointly remove neighbor messages and graph regularization, or fit expression and kinetics in separate stages. Separate fitting preserves each module's update count and uses 5,100 total updates versus 2,700 jointly. WT2 uses the archived all-cell expression fit; U2OS pools three plate-held-out S-only folds within each seed. Velocity metrics use RNA-PCA neighborhoods. The standard deviations measure optimization variability.}
\label{fig:ablation}
\end{figure}

Matched ablations isolated three components of the periodic model: the U-dynamics residual, neighborhood information, and feedback between expression fitting and kinetic fitting. Each condition used the same architecture settings, roots, input partitions and three seeds, with the specified component changed and the model retrained (Fig.~\ref{fig:ablation}; Supplementary Table S11). Removing U dynamics retained U reconstruction and the S equation; removing neighborhoods replaced both neighbor-message operations with self features and removed the geometry penalty. Separate fitting retained the update budget for each module while fitting the expression trajectory before the kinetic parameters.

Removing the U-dynamics residual increased phase error by 0.329 percentage points in WT2 and 0.404 in U2OS. Removing neighborhood information increased it by 0.219 and 0.060 percentage points, respectively. Separate fitting increased it by 0.692 and 0.236 percentage points. Every paired seed showed the same direction of phase-error change for these comparisons. All three ablations also reduced neighbor velocity coherence in both datasets. For example, removing U dynamics reduced the mean coherence from 0.916 to 0.864 in WT2 and from 0.946 to 0.867 in U2OS.

Forward-reference cosine showed a different pattern: several ablations increased this metric despite worsening phase error and local coherence. The full model combined the lowest phase error with the highest local coherence across these matched conditions. The forward-reference endpoint captures a different property of the velocity field and is reported alongside them.

\FloatBarrier

\subsection{Pancreatic differentiation connects lineage progression with RNA dynamics}
\begin{figure}[!p]
\centering
\includegraphics[width=\textwidth,height=.73\textheight,keepaspectratio]{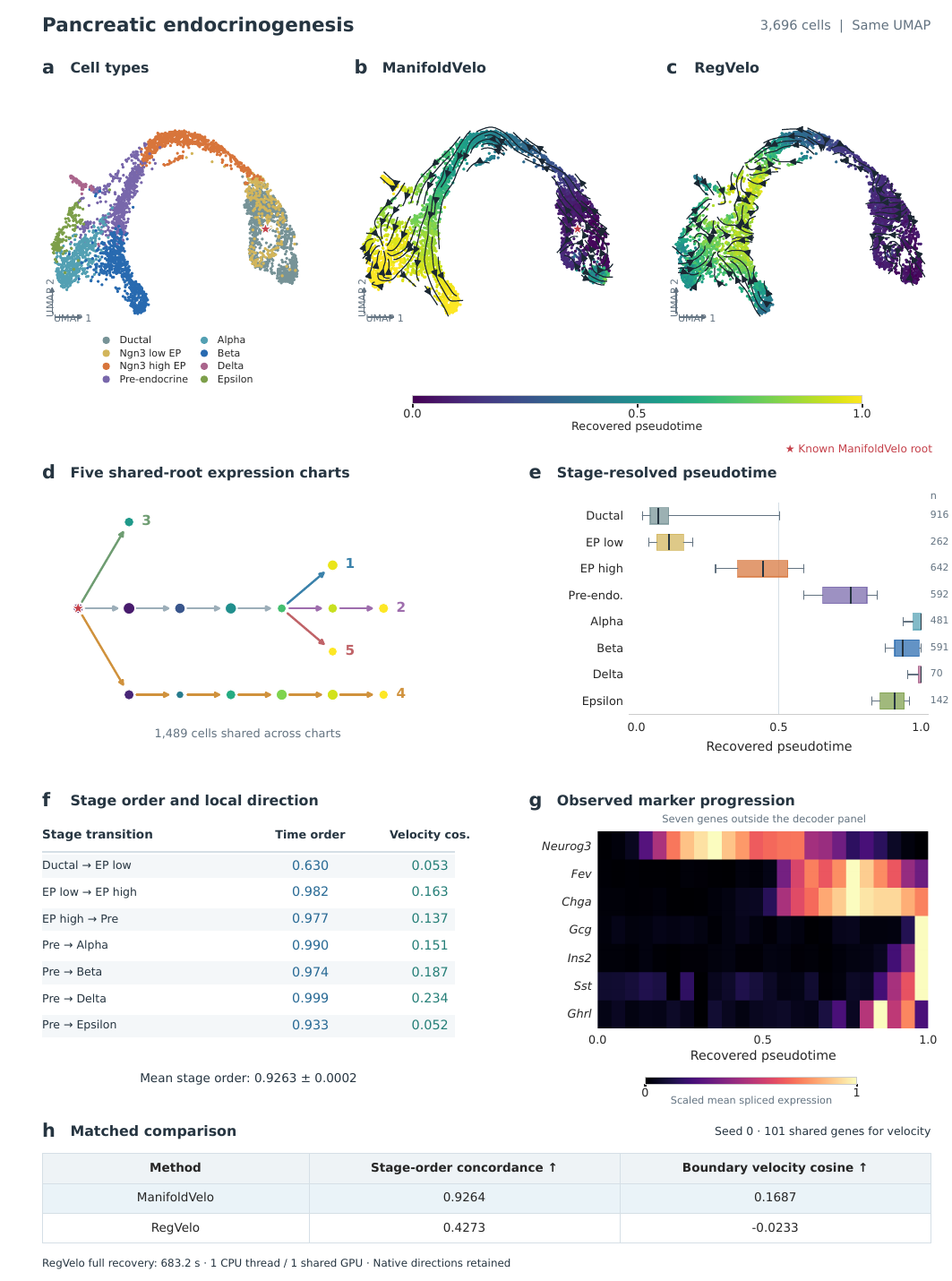}
\caption{\textbf{Pancreatic endocrine progression, lineage direction and marker expression.}
\textbf{a}, Source cell types. \textbf{b,c}, ManifoldVelo and RegVelo time with native S-velocity streamlines on the same UMAP; all 3,696 cells receive time. Stars identify the supplied ManifoldVelo root.
\textbf{d}, Five archived expression-graph paths, sharing 1,489 cells. Nodes are colored by median ManifoldVelo time; arrows show graph orientation.
\textbf{e}, ManifoldVelo time by cell type: median, interquartile range and 10th--90th percentile whiskers; counts at right.
\textbf{f}, Original ManifoldVelo stage-order probabilities averaged across three seeds, and seed-0 velocity cosines using the full gene panel. EP, endocrine progenitor; Pre, pre-endocrine.
\textbf{g}, Seven observed markers outside the decoder panel, averaged over 24 pseudotime bins and scaled within gene. These markers were present in the original expression-graph panel.
\textbf{h}, Matched seed-0 comparison of stage-order concordance and velocity direction over seven transitions. Velocity scoring uses shared retained genes and the same expression-neighbor graph; both methods exclude the supplied root from scoring. This shared-gene endpoint differs from panel f. Velocities are scored before display smoothing.}
\label{fig:pancreas_detail}
\end{figure}

To examine a real branching process in detail, we analyzed 3,696 pancreatic cells spanning ductal states, endocrine progenitors and four hormone-expressing cell types from the study of Bastidas-Ponce and colleagues\cite{pancreas2019}. The frozen model used 149 genes, a supplied ductal root and five expression-derived lineage charts, with 1,489 cells shared across charts. Cell annotations, recovered time and the native kinetic field were displayed on the original UMAP (Fig.~\ref{fig:pancreas_detail}a,b,d). The five charts describe expression-graph paths, while the original cell-type labels provide the biological interpretation of the recovered progression.

Median pseudotime increased from 0.076 in ductal cells to 0.113 in Ngn3-low progenitors, 0.446 in Ngn3-high precursors and 0.754 in pre-endocrine cells; the four hormone-expressing populations had medians between 0.907 and 1.000 (Fig.~\ref{fig:pancreas_detail}e). Across three optimization seeds, the seven-transition stage-order score was $0.9263\pm0.0002$ (Table~\ref{tab:open}). Transition-level concordance was 0.630 for ductal to Ngn3-low cells and 0.933--0.999 for the six subsequent transitions. A complementary check used the native S-velocity vectors at expression-neighborhood boundaries between annotated stages. Mean directional cosines were positive for all seven transitions, ranging from 0.052 to 0.234 (Fig.~\ref{fig:pancreas_detail}f). These readouts connect the recovered temporal order to local motion toward later annotated states in gene space.

Observed marker expression provided a molecular view of this progression (Fig.~\ref{fig:pancreas_detail}g). The binned expression maximum of \textit{Neurog3} occurred at pseudotime 0.35, followed by \textit{Fev} at 0.77. \textit{Chga} increased toward the pre-endocrine region, and hormone-gene expression appeared later along the trajectory. Across the original cell annotations, mean \textit{Gcg}, \textit{Ins2}, \textit{Sst} and \textit{Ghrl} expression was highest in alpha, beta, delta and epsilon cells, respectively. All seven markers were outside the 149-gene decoder panel. Their observed expression and the source stage annotations support biological concordance of the fitted trajectory without using the earlier TSvelo ordering as the reference.

The same shared-root representation was also applied to dentate gyrus, LARRY and erythroid data (Table~\ref{tab:open}). In dentate gyrus, time had a Spearman correlation of $0.9728\pm0.0007$ with the designated earlier TSvelo ordering, with correlations of approximately 0.9929 and 0.9823 within the two fixed exclusive branches. LARRY and erythroid time estimates correlated with sampling day at approximately 0.3193 and 0.8640, respectively. These endpoints complement the stage- and marker-based analysis in pancreas across additional developmental systems.

\begin{table}[!htbp]
\centering\small
\caption{\textbf{Open-lineage applications of the fixed shared-root manifold.}
Values are means $\pm$ sample standard deviations over three optimization seeds.}
\label{tab:open}
\begin{tabular}{@{}llr@{}}
\toprule
Dataset & Reference endpoint & Score \\
\midrule
Dentate gyrus & TSvelo-order Spearman $\rho$ & 0.9728 $\pm$ 0.0007 \\
Pancreas & Coarse-stage precedence & 0.9263 $\pm$ 0.0002 \\
LARRY & Sampling-day Spearman $\rho$ & 0.3193 $\pm$ 0.0002 \\
Erythroid & Sampling-day Spearman $\rho$ & 0.8640 $\pm$ 0.0004 \\
\bottomrule
\end{tabular}
\par\smallskip\footnotesize
Dentate uses the designated earlier TSvelo ordering. Pancreas uses coarse stage order; LARRY and erythroid use sampling day. These endpoints measure different aspects of temporal recovery and are not pooled into a common ranking. All four use the archived known roots and expression-derived charts.
\end{table}

\FloatBarrier

\clearpage
\begingroup
\widowpenalty=10000
\clubpenalty=10000
\section{Discussion}
ManifoldVelo connects temporal ordering and RNA velocity through a continuous expression model whose geometry and kinetics are fitted together. The recovery of complex cyclic and branching simulations shows the value of estimating both a cell's position and its direction of expression change within this representation. A central feature is that the kinetic constraint acts on the derivatives of both reconstructed RNA species, including at coordinates between observed cells. This makes local dynamics informative for fitting the trajectory, while the trajectory provides a continuous reference against which the kinetic field can be evaluated.

VeloCycle established a probabilistic framework for manifold-constrained RNA velocity, linking a spliced-expression manifold to expected unspliced counts and a latent velocity function; its published extensions also cover nonperiodic geometries\cite{velocycle2024}. ManifoldVelo contributes a specific coupling of explicit U/S expression decoders, regulatory transcription--splicing--degradation kinetics and collocation residuals for both RNA derivatives. Periodic neural decoders accommodate the multipeak expression curves examined here, while shared-root lineage decoders represent multiple terminal paths with overlapping progenitor populations. Cellular coordinates, expression functions and kinetic parameters participate in the same fitting objective. These choices allow periodic closure and lineage continuity to enter the expression model directly, connecting structural assumptions to the local RNA changes used to orient progression.

The simulations and component ablations provide complementary evidence for this formulation. In the two multipeak cycles and two branching systems, mean query S-velocity cosines exceeded 0.94, demonstrating accurate gene-space directions when query cells supplied only S. The paired ablations in WT2 and U2OS showed that retaining U dynamics, neighborhood information and joint fitting improved phase recovery and local velocity coherence. Forward-reference cosine responded differently to several ablations, emphasizing that temporal accuracy, local coherence and reference-direction agreement capture distinct properties of an inferred process. Evaluating these properties together helps determine whether a recovered trajectory supports both cellular ordering and local expression change.

The real-data results support applications to cell-cycle phase inference and developmental progression. Under the frozen comparison, phase error decreased relative to the recorded VeloCycle results by 47.0\% in WT2 and 8.7\% in plate-held-out U2OS. The RNA-derived WT2 reference and the FUCCI measurements in U2OS provide different forms of temporal evidence, with the latter supporting evaluation against an independently measured reporter. VeloCycle performed better on WT1 and the smooth cyclic simulation, indicating that the benefit depends on the dataset and trajectory representation. The open-lineage applications extend the same fitting principle to developmental systems, where the available references assess lineage ordering, stage precedence or association with sampling day.

A fitted manifold also has practical value as a reusable dynamic reference. The S-only experiments show that cells can receive temporal coordinates, reconstructed U/S states and velocities without updating the reference parameters. In U2OS, leaving out an entire plate tests this operation across a sample partition; in simulations, the withheld U and known velocity permit direct evaluation of predictions. This capability could support reference-based annotation of additional samples when their cellular states are represented in the fitted trajectory. It also gives a concrete target for future studies of transfer across donors or experimental conditions, where reference coverage and kinetic changes can be evaluated separately.

Further development can focus on selecting the trajectory representation and strengthening independent temporal validation. The present benchmarks specify the cyclic or open family, and the real-data analyses use predefined anchors; the automatic-anchor results establish feasibility in the simulated systems. Selecting topology and anchors from RNA together with progenitor-state information is therefore a useful next step. Donor-level validation with model settings fixed in advance would test transfer beyond the development datasets, while lineage tracing or longitudinal measurements could assess branch direction and progression more directly. The current coordinate describes normalized trajectory time; independent temporal calibration would support estimates of absolute transition rates. These extensions would build on the explicit correspondence between temporal position, RNA expression and local dynamics established by the present model.

\subsection*{Conclusion}
Single-cell RNA snapshots leave both the order of cellular states and the direction of change unresolved. Similar expression patterns may recur during a cycle, while cells with a common progenitor state may enter different lineages. ManifoldVelo fits continuous unspliced and spliced RNA trajectories, estimating cell positions and kinetic parameters together. The kinetic equations constrain the slopes of both expression curves at sampled cells and between them. RNA dynamics thus helps refine cell order, while the trajectory constrains the inferred velocity field.

The model describes cycles with closed curves and branching lineages with curves that share a common origin. An open curve represents unbranched progression, and neighborhood relationships guide the local fit in each case. These geometric choices allow the same kinetic formulation to be used across the cyclic, branching and smooth datasets examined here. The resulting trajectories recover temporal progression together with local RNA dynamics. Once fitted, they also provide a reference for assigning pseudotime and predicting velocity in additional cells measured with spliced RNA alone. This design makes it possible to reuse learned RNA dynamics for new cells, establishing ManifoldVelo as a general framework for joint pseudotime and velocity inference across cyclic, branching and unbranched trajectories.

\par\endgroup
\section{Methods}
\subsection{Model definition and implementation}
Let a temporal domain $\mathcal Q$ index a differentiable RNA expression function $Y:\mathcal Q\to\R_{\geq0}^{2G}$. ManifoldVelo estimates a coordinate $q_i$ for each reference cell, the parameters of $Y$, and a regulatory kinetic field $f$. RNA reconstruction places observed cells on the expression function, while a collocation residual constrains $\dd Y/\dd q$ to agree with $f(Y(q),q)$. The temporal domain is a circle for a cyclic process or a collection of root-to-terminal intervals for an open process. Open intervals have a common decoded root and overlapping membership for shared progenitors. A single interval represents an unbranched process.

Topology conditioning specifies a coordinate parameterization, an expression decoder and its structural constraints. The recorded experiments use fixed topology families and dataset-specific presets. Within the open family, expression-derived lineage paths determine the number of decoders. The periodic WT2/U2OS configuration uses a graph encoder and Fourier decoder; multipeak simulations use an order-preserving circular map and a periodic neural decoder; open trajectories use lineage-specific neural curves. These are alternative active configurations of the framework, as summarized in Fig.~\ref{fig:framework}b and Table~\ref{tab:architecture}.

The software was developed from the TSvelo implementation\cite{tsvelo2026}. Its regulatory transcription--splicing--degradation component provides the kinetic foundation; ManifoldVelo adds explicit continuous U/S decoders, topology-dependent coordinate representations, tangent--kinetic residuals and fixed-reference query procedures. The evaluated core is the release frozen on 16 September 2026, together with the subsequent automatic-root preprocessing wrapper for simulations. Reference times are used for calibration and evaluation. Archived preset selection was guided by reference-label evaluations during development, whereas the training loss and S-only query input exclude those labels. The inferred coordinate is normalized trajectory time rather than an independently calibrated duration in hours.

\subsection{RNA inputs, masks and the reference expression graph}
Let $U,S\in\R_{\geq0}^{N\times G}$ denote the prepared unspliced and spliced RNA matrices for $N$ reference cells and $G$ model genes. The observed RNA state is $y_i=(U_i,S_i)\in\R^{2G}$. A channel-observation mask $m_{ia}$ specifies whether channel $a$ is measured for cell $i$. Selected paired genes contribute both U and S, whereas auxiliary S-only genes contribute only S. We write $\activeset$ for the set of active RNA channels and $a_{\rm act}\in\{0,1\}^{2G}$ for its indicator. This observation mask is distinct from the regulator mask and lineage-membership matrix introduced below.

For the periodic graph encoder, gene features are
\begin{equation}
x_{ig}=\operatorname{clip}\left(
\frac{\log(1+S_{ig})-\mu_g}{\max(\sigma_g,0.1)},-8,8\right),
\label{eq:features}
\end{equation}
where the mean and standard deviation are computed from reference cells. Principal components of $x$ define neighborhood distances; the encoder itself receives the full $G$-gene feature vector. WT2/U2OS use up to 20 principal components and 30 nearest reference neighbors. The multipeak chart uses 10 components and 10 neighbors. Dimensions and neighbor counts are capped by the available data size. Self edges are excluded. If $d_{ij}$ is a retained neighbor distance and $\sigma_d$ is the median positive retained distance, graph weights are
\begin{equation}
A_{ij}=\exp[-d_{ij}^{2}/(2\sigma_d^{2})],\qquad
A\leftarrow\max(A,A^{\mathsf T}).
\label{eq:graph}
\end{equation}
Numerical edge weights are floored at $10^{-8}$. Weighted spectral initialization uses the second and third eigenvectors of the normalized graph Laplacian, standardized and normalized to a circle. GraphSAGE aggregation and the geometric loss use the unweighted adjacency of this graph. WT2/U2OS instead initialize the circular angle from the first two standardized principal components. RNA residuals use a separate fixed scale $r_a=\max[\operatorname{sd}(y_{\cdot a}),0.1]$ in the original prepared RNA units.

\subsection{Circular coordinates and the graph neural encoder}
The cyclic coordinate is a phase $p_i\in[0,1)$ with circular embedding
\begin{equation}
z(p_i)=(\sin 2\pi p_i,\cos 2\pi p_i).
\label{eq:circle}
\end{equation}
The encoder contains two mean-aggregation GraphSAGE layers\cite{graphsage2017}. With $h_i^0=x_i$, layer $\ell\in\{0,1\}$ computes
\begin{equation}
h_i^{\ell+1}=\operatorname{SiLU}\!\left(
A_\ell h_i^\ell+B_\ell\frac{1}{|\mathcal N_i|}\sum_{j\in\mathcal N_i}h_j^\ell+b_\ell\right).
\label{eq:sage}
\end{equation}
Both hidden representations have dimension 128. Separate root and neighbor transformations are learned; the root transformation is bias-free and the neighbor transformation supplies the bias shown in Eq.~\eqref{eq:sage}. An affine $128\to2$ head gives $\widetilde z_i$, which is normalized to unit length. Phase is recovered using $\operatorname{atan2}(z_{i1},z_{i2})/(2\pi)$ modulo one. Numerical guards handle near-zero outputs. Thus, the active encoder has two message-passing layers and one output head. It contains no dropout or batch normalization.

Initial orientation is selected using training U/S. A second-harmonic least-squares curve through the initial S coordinates supplies an expression tangent. Its scale-adjusted inner product with a provisional field $\beta_0U-\gamma_0S$ chooses the global direction, with $\beta_0=2\pi$ and $\gamma_{0g}=\operatorname{clip}(2\pi\overline U_g/\max[\overline S_g,0.01],0.1,100)$. The selected root shifts the coordinate to the configured phase $p_\star=0.05$. This value is an arbitrary origin convention. A ridge-initialized output head starts the learned encoder near the oriented chart.

\subsection{Periodic RNA decoders and order-preserving circular maps}
WT2 and U2OS use a second-order Fourier decoder. The deterministic feature map
\begin{equation}
F_2(p)=(\sin 2\pi p,\cos 2\pi p,\sin 4\pi p,\cos 4\pi p)
\end{equation}
is followed by one affine $4\to2G$ readout. The decoded state is
\begin{equation}
Y(p)=a_{\rm act}\odot r\odot
\exp\!\left[\operatorname{clip}(C F_2(p)+c,-16,16)\right].
\label{eq:fourier}
\end{equation}
This decoder has no hidden neural layer. The coordinate dependence guarantees periodic closure. A more flexible alternative uses a multilayer perceptron (MLP),
\begin{equation}
Y(p)=a_{\rm act}\odot r\odot\spplus\!\left(
C_3\tanh[C_2\tanh(C_1z(p)+c_1)+c_2]+c_3\right),
\label{eq:periodic_mlp}
\end{equation}
with widths $2\to H\to H\to2G$. There are two Tanh hidden layers and one affine output layer. The multipeak data1/data2 configurations use $H=128$; the smooth-cycle and RPE1 configurations use $H=64$.

The multipeak model learns phase by warping a reference circular ordering. Adaptive knots $0=\kappa_0<\cdots<\kappa_K=1$ are obtained from every fourth sorted reference phase and the endpoints. Let $w_k=\kappa_{k+1}-\kappa_k$ and let $a_k$ be trainable log multipliers. Positive interval increments are
\begin{equation}
\delta_k=\frac{w_k\exp[\operatorname{clip}(a_k,-12,12)]}
{\sum_j w_j\exp[\operatorname{clip}(a_j,-12,12)]},\qquad
\widetilde\delta_k=\frac{\delta_k+10^{-7}}{1+K10^{-7}}.
\end{equation}
For a reference phase $p\in[\kappa_k,\kappa_{k+1})$, the learned map is
\begin{equation}
T(p)=\left[p_\star+\sum_{j<k}\widetilde\delta_j+
\frac{p-\kappa_k}{w_k}\widetilde\delta_k\right]\bmod1.
\label{eq:warp}
\end{equation}
The base chart is shifted so that its root is zero, and the offset $p_\star$ is fixed. Positive increments preserve the reference circular ordering while learning nonuniform temporal spacing. Adjacent log multipliers receive a weak cyclic smoothness penalty. The retained GraphSAGE module is frozen and bypassed in this route; the active trained coordinate is Eq.~\eqref{eq:warp}. RNA and kinetic parameters are fitted jointly with the interval multipliers.

\subsection{Open lineage charts and the shared-root neural decoder}
Open trajectories are initialized using diffusion pseudotime (DPT)\cite{dpt2016} and partition-based graph abstraction (PAGA)\cite{paga2019} in Scanpy\cite{scanpy2018}. Neighbors are formed from the supplied S-derived representation, Leiden clusters define coarse states, and PAGA supplies their connectivity. The configured root orients a maximum spanning tree of the positive PAGA connections in the standard construction. Maximal root-to-terminal paths define $B$ charts. A cell belongs to every chart containing its coarse state, so common-trunk cells can appear in several charts. Let $M_{ib}\in\{0,1\}$ denote this membership. The root belongs to every chart. DPT ranks, normalized to $[0,1]$, supply initial time $t_i^0$ with the root fixed at zero.

Each lineage uses an independent $1\to96\to96\to2G$ MLP $g_b$, with two Tanh hidden layers and an affine readout. The RNA decoder is
\begin{equation}
Y_b(t)=a_{\rm act}\odot r\odot\spplus[a_0+t\,g_b(t)],\qquad 0\leq t\leq1.
\label{eq:open}
\end{equation}
The trainable $2G$-vector $a_0$ is shared by all lineages and initialized from the earliest ten DPT cells. Consequently, $Y_b(0)$ is exactly equal across lineages. Interior shared-trunk agreement is learned from overlapping observations and the penalties below. A single chart gives a conventional open trajectory with the same implementation. Network weights in $g_b$ are lineage-specific; the root vector and regulatory matrix are shared.

During warm-up, a non-neural 48-bin monotone clock $h_b(t_i^0)$ adjusts temporal spacing. Positive normalized increments, formed from the softmax of trainable log-bin values with a small positive floor, define a piecewise-linear map with endpoints zero and one. After warm-up, clock parameters are frozen and individual cell times $t_{ib}$ are updated by curve projection. These later updates can change within-chart ordering subject to the reconstruction and anchoring objective. The warm-up map is therefore an initialization of open pseudotime, distinct from a cyclic physical-clock factor.

\begin{table}[!htbp]
\centering\small
\caption{\textbf{Active network architectures in the evaluated configurations.} Width sequences include input, hidden and output dimensions. ``Paired'' counts genes with active U and S channels; all listed model genes have S channels.}
\label{tab:architecture}
\setlength{\tabcolsep}{4pt}
\begin{tabularx}{\textwidth}{@{}lrrXl@{}}
\toprule
Dataset & $G$ & Paired & Coordinate module & RNA decoder \\
\midrule
WT1 & 375 & 375 & GraphSAGE 128, 128 & Fourier $4\to750$ \\
WT2 & 391 & 391 & GraphSAGE 128, 128 & Fourier $4\to782$ \\
U2OS (each fold) & 107 & 107 & GraphSAGE 128, 128 & Fourier $4\to214$ \\
Multipeak data1/2 & 18 & 18 & Monotone circular map & $2\to128\to128\to36$ \\
Smooth cycle & 18 & 18 & GraphSAGE 128, 128 & $2\to64\to64\to36$ \\
RPE1 & 28 & 20 & GraphSAGE 128, 128 & $2\to64\to64\to56$ \\
Linear simulation & 36 & 36 & One open chart & $1\to96\to96\to72$ \\
Trifurcation & 36 & 36 & Three open charts & $1\to96\to96\to72$ \\
Nested branches & 36 & 36 & Four open charts & $1\to96\to96\to72$ \\
Dentate gyrus & 165 & 100 & Two open charts & $1\to96\to96\to330$ \\
Pancreas & 149 & 100 & Five open charts & $1\to96\to96\to298$ \\
LARRY & 591 & 591 & Three open charts & $1\to96\to96\to1182$ \\
Erythroid & 591 & 591 & Five open charts & $1\to96\to96\to1182$ \\
\bottomrule
\end{tabularx}
\par\smallskip\footnotesize
Each open chart has its own two-hidden-layer decoder. The only additional active residual MLP is $2\to64\to18\to18$ in the multipeak simulations. The optional cyclic clock MLP in the source is disabled in all configurations listed here.
\end{table}

\subsection{Regulatory transcription and coupled U/S kinetics}
The kinetic component derives from the TSvelo regulatory transcription--splicing--degradation formulation\cite{tsvelo2026}. A ReLU transcription link couples regulatory inputs to RNA production; continuous U/S decoders connect this field to the learned trajectory through collocation residuals. For a paired gene $g$, the kinetic field is
\begin{gather}
\begin{gathered}
\frac{\dd U_g}{\dd\tau}=\alpha_g(S,q)-\beta_gU_g,\\[3pt]
\frac{\dd S_g}{\dd\tau}=\beta_gU_g-\gamma_gS_g,
\end{gathered}
\label{eq:ode}\\[4pt]
\alpha(S,q)=\relu\!\left(W_{\rm eff}S+b+C_{\rm drive}\psi(q)+r_{\rm res}(q)\right).
\label{eq:alpha}
\end{gather}
Here $q$ is cyclic phase or open time, $W_{\rm eff}$ represents cross-gene transcriptional coupling, $b$ is a transcription bias, and $\psi$ is a coordinate-dependent drive. The cyclic drive uses $(\sin2\pi p,\cos2\pi p)$; the open drive uses $(t,t^2,t^3)$. With a forbidden-edge indicator $F$, the effective regulator matrix is $W_{\rm eff}=W\odot(1-F)$. A permitted element $W_{gh}$ describes input from gene $h$ to transcription of gene $g$. Its learned sign can represent activation or repression before the nonnegative transcription link. The reference WT2 mask permits E2f1 and E2f6 as candidate regulator columns; the U2OS mask permits E2F1--E2F8, FOXM1 and MYBL2. These are candidate-TF masks, not curated TF--target edge matrices. The reference fits do not add a curated edge-specific penalty.

Splicing and degradation rates are positive, with $\beta_g=\spplus(\widetilde\beta_g)+\epsilon$ and $\gamma_g=\spplus(\widetilde\gamma_g)+\epsilon$. The floors are $10^{-8}$ for the cyclic backend and $10^{-6}$ for the open backend. Open models have lineage-specific rates, biases and polynomial drives, while sharing $W$. Auxiliary S-only coordinates use $\dd S_g/\dd\tau=\alpha_g-\gamma_gS_g$, with their U state and U derivative masked to zero. This preserves their regulatory contribution without treating unobserved U as measured zeros.

In multipeak models, the residual transcription term has a $2\to64\to18\to G$ MLP, two Tanh hidden activations and a bias-free final readout. An outer Tanh bounds the result:
\begin{equation}
r_{\rm res}(p)=8\,s_\alpha\odot\tanh[g_{\rm res}(z(p))],
\end{equation}
where $s_\alpha$ is the fixed gene-wise transcription scale. The residual is disabled in the WT2/U2OS and open configurations. The GRN term itself is a masked learned matrix; it is not another multilayer network.

\subsection{RNA reconstruction and tangent--kinetic consistency}
Let $Y_i$ denote the decoded RNA state at a cell's current coordinate. The cyclic reconstruction term gives equal weight to active RNA channels after scaling,
\begin{equation}
\mathcal L_{\rm RNA}=\frac{1}{|\activeset|}\sum_{a\in\activeset}
\frac{\sum_{i\in\mathcal B}m_{ia}[(Y_{ia}-y_{ia})/r_a]^2}
{\sum_{i\in\mathcal B}m_{ia}},
\label{eq:recon}
\end{equation}
where $\mathcal B$ is the current cell batch and channels without batch observations are excluded. WT2/U2OS and the reported simulations reconstruct prepared RNA values with this squared-error objective. WT1 uses its archived log-reconstruction option; RPE1 uses its archived robust reconstruction option. Input log-S transformation for graph construction is separate from the RNA-space reconstruction scale.

Automatic differentiation evaluates the decoder tangent $Y'(q)$. At randomly offset, uniformly spaced collocation coordinates $q_c$, the dynamic constraint is
\begin{equation}
\mathcal L_{\rm dyn}=\frac{1}{C|\activeset|}\sum_{c=1}^{C}\sum_{a\in\activeset}
\left[\frac{Y'_a(q_c)/\rho(q_c)-f_a(Y(q_c),q_c)}{r_a}\right]^2,
\label{eq:dyn}
\end{equation}
where $f$ is Eq.~\eqref{eq:ode} and $\rho=\dd\tau/\dd q$ denotes a possible coordinate-to-clock factor. All fixed cyclic configurations analyzed here have $\rho=1$; open tangents are also evaluated directly with respect to their normalized model time. Collocation ties the entire learned RNA curve, including intervals between observed cells, to the kinetic field. The curve is fitted under these residual constraints rather than obtained by exact numerical integration of the ODE. Both $f(Y(q),q)$ and $Y'(q)$ are retained in the output.

Curve curvature is regularized through finite differences of neighboring tangents, scaled by $C$ and the RNA scale. Cyclic differences wrap from the last collocation point to the first; open differences use only interior neighbors. An optional angular term averages $1-\cos(D^{-1}Y',D^{-1}f)$, where $D=\operatorname{diag}(r)$. It is active in the smooth-cycle configuration with weight 0.03 and is zero in the WT2/U2OS reference configuration. Thus, the WT2/U2OS objective couples tangent direction and magnitude through Eq.~\eqref{eq:dyn} itself.

\subsection{Geometric constraints and the cyclic objective}
Define $\Delta(p,p')=\wrap(p-p')\in[-1/2,1/2)$ as the shortest signed circular difference. For observed graph edges $\mathcal E$ and sampled nonedges $\mathcal E^{-}$, the geometry term is
\begin{align}
\mathcal L_{\rm geom}={}&\cmean_{(i,j)\in\mathcal E}[2\pi\Delta(p_i,p_j)]^2\nonumber\\
&+\cmean_{(i,j)\in\mathcal E^{-}}\relu[1-|2\pi\Delta(p_i,p_j)|]^2.
\label{eq:geometryloss}
\end{align}
The first term retains neighborhood proximity; the second discourages collapse of separated cells. Nonedges exclude self pairs and existing graph edges. Initialization anchoring, radius stabilization and root anchoring are
\begin{align}
\mathcal L_{\rm init}&=\cmean_i[2\pi\Delta(p_i,p_i^0)]^2,\\
\mathcal L_{\rm radius}&=\cmean_i(\|\widetilde z_i\|_2^2-1)^2,\\
\mathcal L_{\rm root}&=\cmean_{i\in\mathcal R}[1-\cos 2\pi(p_i-p_\star)],
\end{align}
where $\mathcal R$ contains the configured root cells. The WT2/U2OS joint objective is
\begin{align}
\mathcal L_{\rm cycle}={}&\mathcal L_{\rm RNA}
+0.1\mathcal L_{\rm dyn}+0.05\mathcal L_{\rm geom}
+0.02\mathcal L_{\rm init}\nonumber\\
&+0.001\mathcal L_{\rm radius}+0.1\mathcal L_{\rm root}
+10^{-5}\mathcal L_{\rm curv}+0.001\mathcal L_{\rm rate}.
\label{eq:cyclictotal}
\end{align}
The rate penalty is the sum of mean squared log-rate ratios to their post-warm-up initial values, mean squared allowed regulator weights, and mean squared transcription-drive coefficients divided by $s_\alpha$. Warm-up uses reconstruction and the geometry, radius, root and initialization terms, with initialization weight 1.0. The multipeak route replaces the learned graph coordinate with the monotone map and uses its archived reconstruction, circular-anchor, interval-smoothness, dynamics, residual, curvature and rate penalties. The root is enforced by its fixed map offset.

\subsection{Open-trajectory objective and alternating time projection}
Open reconstruction and dynamics are averaged equally across lineage charts. Within chart $b$, each term is evaluated on its member cells or collocation points using chart-specific channel scales $r_{ba}$, floored at 0.1. Let $\mathcal B_i=\{b:M_{ib}=1\}$, and let $\overline Y_i$ be the mean of decoded chart states for cell $i$. Shared-state overlap is penalized by
\begin{equation}
\mathcal L_{\rm overlap}=\cmean_{i:|\mathcal B_i|>1}
\frac{1}{|\mathcal B_i|}\sum_{b\in\mathcal B_i}
\|D_0^{-1}[Y_b(t_{ib})-\overline Y_i]\|_{\activeset}^2,
\label{eq:overlap}
\end{equation}
where $D_0$ uses the global training RNA scale and $\|v\|_{\activeset}^2$ denotes the mean squared active-channel value. An anchor penalizes $\cmean_{i,b:M_{ib}=1}(t_{ib}-t_i^0)^2$, and adjacent log-bin clock parameters receive a squared-difference penalty. During joint fitting, dynamic residuals, curvature and rate penalties are added. The open rate term penalizes allowed regulator weights and scaled drive coefficients, with a weaker squared-log penalty on positive rates. Optional shared-velocity agreement applies only to cells observed in multiple charts; it does not force daughter lineages to point in the same direction after separation.

After clock warm-up, curve/kinetic optimization alternates with discrete projection of cell times. Every 200 joint updates in the reported open simulation protocol, candidate times are evaluated on a 1,000-point grid:
\begin{equation}
t_{ib}^{\rm new}=\arg\min_{t\in\mathcal T}
\left\{\cmean_{a\in\activeset}\left[\frac{Y_{ba}(t)-y_{ia}}{r_{ba}}\right]^2
+\lambda_{\rm anchor}(t-t_i^0)^2\right\}.
\label{eq:projection}
\end{equation}
An update is accepted only if it reduces the corresponding cost; the root remains at zero and chart membership remains fixed. Configuration-specific shared-time or movement terms can be added to this cost; their recorded simulation settings leave those extra restrictions inactive. Model fitting therefore updates the expression curves, kinetic parameters and cell times while retaining the lineage organization.

For a cell appearing in several charts, the exported time, state and velocity are separately averaged using
\begin{equation}
\omega_{ib}=\frac{M_{ib}n_b}{\sum_c M_{ic}n_c},\qquad
n_b=\sum_i M_{ib}.
\label{eq:membershipweight}
\end{equation}
Per-chart coordinates and membership are also retained. An unused cell--chart pair has no chart-specific time; every modeled cell still receives a global output from its active chart memberships.

\subsection{Initialization, training schedules and stopping rules}
Fitting begins with reconstruction warm-up. Positive kinetic rates are then initialized from the learned RNA curve: bounded least squares fits $[U,-S](\beta_g,\gamma_g)^{\mathsf T}$ to the S tangent, with rates constrained to $[10^{-4},100]$. A nonnegative transcription target is obtained from $Y'_U+\beta U$, or $Y'_S+\gamma S$ for auxiliary S-only coordinates. Conditional ridge fitting initializes the transcription bias and coordinate drive. The cyclic design uses $[1,\sin2\pi p,\cos2\pi p]$; the open design uses $[1,t,t^2,t^3]$. The nonconstant drive coefficients have ridge weight 0.01. Joint optimization then allows reconstruction and kinetic consistency to update the expression curve together.

WT2/U2OS use Adam with learning rate $10^{-3}$, weight decay $10^{-5}$, cell batch size 256, gradient-norm clipping at 5, 300 warm-up updates and 2,400 joint updates. The reference graph is encoded as a whole at each step; reconstruction uses sampled cells and dynamics uses 64 collocation points. The open simulation schedule uses 800 warm-up and 3,200 joint updates, batch size 2,048, 96 collocation points, and Adam with cosine learning-rate decay from 0.002 to 0.0001. Trifurcation/nested-branch dynamics weights are 0.03; the linear case uses 0.01. Their anchor, overlap, clock-smoothness, curvature and rate weights are 0.2, 0.1, 0.001, $10^{-5}$ and $10^{-4}$, respectively.

Multipeak models use separate learning rates of 0.002 for model parameters and 0.004 for circular-map parameters, weight decay $10^{-6}$ and 192 collocation points. Their dynamics and circular-anchor weights are both $10^{-3}$; interval smoothness, residual and rate weights are $10^{-5}$, and curvature weight is $10^{-8}$. Following initial reconstruction fitting, kinetic initialization occurs at step 1,501 and dynamic terms become active. The maximum is 18,000 updates; stability is evaluated every 500 updates after the minimum of 5,000. Three consecutive passes are required for the recorded objective, phase, curve and parameter-change tolerances of $2\times10^{-4}$, $5\times10^{-4}$, $3\times10^{-3}$ and $10^{-2}$, respectively. Data1 and data2 stopped at 13,500 and 15,000 updates in the complete automatic-root replay. Other presets use their stated fixed budgets. Supplementary Tables S2 and S3 list the network dimensions, optimization settings and principal loss weights for each preset.

\subsection{Automatic root candidates from training U/S}
All six simulation replays infer their anchor from training RNA, without supplied roots or marker lists. The selector standardizes log-S with clipping at five standard deviations, forms a 30-neighbor PCA graph and coarsens it using MiniBatch K-means. The number of groups is $\min(60,\max[16,\operatorname{round}(\sqrt N/2)],N)$. Groups smaller than $\max(8,0.001N)$ cells are ineligible as sources.

For paired genes, the upper 5\% of jointly scaled S/U abundance defines an extreme set $E_g$. A provisional slope and residual are
\begin{equation}
\widehat\gamma_g=\frac{\sum_{i\in E_g}S_{ig}U_{ig}}
{\sum_{i\in E_g}S_{ig}^{2}},\qquad
\widehat v_{ig}=\frac{U_{ig}-\widehat\gamma_gS_{ig}}
{\max[\operatorname{sd}(S_{\cdot g}),0.1]}.
\label{eq:rootvelocity}
\end{equation}
Only genes with nonconstant U/S and positive slope contribute. Let $K_{ab}$ be normalized connectivity between expression groups and let $c_{ab}$ be the cosine between the centered group-mean residual and the centered expression displacement from group $a$ to $b$. Directed connectivity is $P^*_{ab}=K_{ab}\exp(4c_{ab})$. A row-normalized random walk with restart probability 0.03 is evaluated on the reversed directed graph and on the undirected reference. Their stationary masses define
\begin{equation}
\operatorname{source}(a)=\log\frac{\pi_{\rm reverse}(a)}{\pi_{\rm undirected}(a)}.
\end{equation}
The highest-scoring eligible group supplies the cell nearest its PCA centroid. Thirty 80\%-gene subsamples record source-selection stability. The cyclic simulations use the same source selector to choose a computational phase anchor; assigning that anchor to a biological M-to-G1 transition would require additional evidence. The primary real-data comparisons retain their recorded known roots. Supplementary Tables S4 and S5 list the automatic simulation anchors and retained real-data roots, respectively.

\subsection{Reference-based prediction from S-only query cells}
Query input consists of S, gene identifiers and cell identifiers. The stored training transformation is reused, with gene order checked and aligned. For the graph encoder, query neighbors are selected only from the reference cells. At layer $\ell$, the query's own representation is combined with the mean of cached reference representations $h_j^\ell$ using Eq.~\eqref{eq:sage}. Reference representations are fixed, and neither query--query edges nor messages back into the reference are required. The predicted phase is passed through the decoder and kinetic field to obtain U/S reconstruction, the curve tangent and native kinetic velocity.

For a circular-map model, the ten nearest reference positions provide a weighted complex mean of the initial circular phases, with weights proportional to $\exp[-d_j^2/d_{10}^2]$. Applying the learned map $T$ produces the query phase. Open models obtain initial time and soft chart support from reference neighbors, using normalized weights $\exp(-d_j/d_k)$. Each supported chart is then searched on the stored 1,000-point grid by replacing the RNA term in Eq.~\eqref{eq:projection} with its S-only counterpart. Soft support and reference chart size determine output weights as in Eq.~\eqref{eq:membershipweight}. Query U and reference time never enter these prediction steps. The reference model is reused without fitting a new decoder for each query set.

\subsection{Datasets, evaluation protocols and velocity metrics}
The six simulated systems comprise multipeak cycles data1 and data2, a smooth cycle, a linear trajectory, a trifurcation and nested branches. Each cyclic system has 320 paired training cells, 160 S-only query cells and 18 genes; each open system has 720 paired training cells, 360 queries and 36 genes. True time and stored simulator RNA derivatives are reserved for evaluation. TSvelo and TFvelo simulation runs use no supplied regulatory-edge priors. Comparator S-only readouts retain their documented native or interpolation-based query interfaces (Supplementary Table S6).

The periodic real-data examples are WT2 mouse neurosphere cells with tricycle annotations\cite{tricycle2022} and U2OS FUCCI cells from GSE146773\cite{mahdessian2021}. WT2 has 7,057 cells and 391 model genes, with 1,411 phase labels for calibration and 5,646 for evaluation. ManifoldVelo retains author-normalized fractional RNA values; the archived VeloCycle fit internally converts its input to integers. U2OS has three plates of 384 cells and 107 model genes. The frozen S-only test trains on 768 paired cells and queries the remaining 384 cells, repeated over all three plates; 1,146 of the 1,152 cells have FUCCI labels. WT1 provides a companion neurosphere comparison. Dentate gyrus\cite{dentate2018}, pancreas, LARRY and erythroid analyses use their archived roots, charts and reference definitions. Supplementary Table S1 summarizes the datasets, input sizes and evaluation references.

The full-S/U velocity comparison fits all cells and uses a separate calibration/evaluation split: 1,411/5,646 WT2 labels and 228/918 U2OS labels. The six U2OS cells without FUCCI labels still receive model predictions. Comparators include VeloCycle\cite{velocycle2024}, TSvelo\cite{tsvelo2026}, scVelo dynamical and stochastic modes\cite{scvelo2020}, cellDancer\cite{celldancer2024}, TFvelo\cite{tfvelo2024}, veloVI\cite{velovi2024} and RegVelo\cite{regvelo2026}; the simulation archive additionally includes UniTVelo\cite{unitvelo2022}. Methods retain their recorded preprocessing, gene panels, priors and budgets. TFvelo's total-RNA velocity is evaluated against the corresponding total-RNA reference in simulations, separately from S velocity. The regulatory-ODE baseline uses the official implementation and an archived species-matched CollecTRI prior\cite{collectri2023} on real data; synthetic comparisons receive no supplied regulatory-edge network. Supplementary Notes S3 and S5 document all comparator readouts, checkpoint handling and the RNA-only parameter-selection controls. Supplementary Tables S7--S10 report the complete benchmark outcomes; Tables S14--S18 record all parameter-check configurations and repeats. Supplementary Tables S19--S21 specify comparator parameters for simulations, full-S/U real data and frozen periodic comparisons, respectively; Supplementary Table S22 records software versions and source evidence.

Circular mean absolute error is
\begin{equation}
\operatorname{cMAE}=\frac{100}{N_{\rm eval}}\sum_{i\in\mathcal I_{\rm eval}}
\left|\Delta(\widehat p_i,p_i^{\rm ref})\right|,
\end{equation}
expressed as a percentage of a cycle. The frozen comparison retains its calibration-only rotation/reflection protocol. The full-S/U comparison uses a calibration-only global rotation and preserves each method's native velocity direction. Open simulated time uses absolute error on $[0,1]$, also multiplied by 100. Direct velocity accuracy averages per-cell cosines between predicted and true high-dimensional S derivatives. Real-data forward cosine measures alignment with reference-forward neighbor displacements; neighbor cosine measures local alignment of predicted velocity directions. The full-S/U comparison uses fixed UMAP-space neighborhoods, while the component ablations use fixed RNA-PCA neighborhoods. These proxy metrics are computed before display smoothing.

UMAP\cite{umap2018} coordinates are held fixed across methods. Local expression-to-embedding Jacobians project the native velocity field for display, with arrow magnitudes and smoothing adjusted for visibility. The graph or chart construction used by the model is separate from this display. A simulation VeloCycle fit on an open trajectory retains its periodic output; that native clock is not assigned a linear-time MAE. Source tables explicitly identify this metric mismatch and incomplete historical timer boundaries.

The detailed simulation figures use the archived predictions and the same fixed embeddings and projected streamline paths. Circular residual plots show $100\Delta(\widehat p_i,p_i^{\rm ref})$ for each S-only query after the archived training-only rotation/reflection calibration. No additional alignment is fitted to query labels, and velocity directions retain their native orientation. For the branching simulations, empirical cumulative distributions summarize the per-cell S-velocity cosines before projection. As in the original evaluator, reference vectors with norm at most $\max(10^{-8},10^{-6}\operatorname{median}_j\|v_j^{\rm ref}\|_2)$ within the query set are excluded; all 360 queries in each displayed branching system pass this criterion. The cosine denominator is bounded below by $10^{-8}$, so a zero predicted vector contributes zero. The two scVelo dynamical time readouts share one velocity field and therefore one cumulative curve. TFvelo's total-RNA score is displayed separately.

\subsection{Pancreatic stage concordance and marker analysis}
The detailed pancreatic analysis uses the frozen \texttt{dynamics\_weak} predictions for seeds 0, 1 and 2. Cell identifiers align the 3,696 predictions to the embryonic-day-15.5 subset of the original study\cite{pancreas2019}, distributed as the \href{https://scvelo.readthedocs.io/en/stable/scvelo.datasets.pancreas.html}{scVelo pancreas dataset}; UMAP coordinates and native kinetic vectors are retained. The supplied root is the ductal cell \texttt{AAGACCTTCTGACCTC}. The 149-gene decoder comprises 100 paired U/S genes and 49 auxiliary S-only genes. The original 2,000-gene expression graph defines five rooted charts with 1,489 shared cells. The model settings were selected during the recorded development comparison; this analysis characterizes the fixed model against the source annotations and observed expression.

For an earlier stage $A$ and a later stage $B$, temporal concordance is
\begin{equation}
C_{A\to B}=\frac{1}{n_A n_B}\sum_{i\in A}\sum_{j\in B}
\left[\mathbf{1}(t_j>t_i)+\frac{1}{2}\mathbf{1}(t_j=t_i)\right].
\label{eq:pancreas_order}
\end{equation}
The seven comparisons are ductal to Ngn3-low, Ngn3-low to Ngn3-high, Ngn3-high to pre-endocrine, and pre-endocrine to each of alpha, beta, delta and epsilon cells. The root is excluded. The aggregate weights all seven comparisons equally. Three-seed means and sample standard deviations reproduce the archived score and summarize optimization variability.

For a local velocity-direction check, we form 30-neighbor neighborhoods from 20 principal components of standardized, clipped log-S features over the 149 model genes, using the feature transformation in Eq.~\eqref{eq:features}. For each source cell in $A$ with neighbors in $B$, we compute the cosine between its native $v_{S,i}$ and each displacement $S_j-S_i$, after dividing both vectors gene-wise by $\max[\operatorname{sd}(S_{\cdot g}),0.1]$. Cosines are averaged over the eligible neighbors and then equally over source cells. The seven transitions have 811, 55, 124, 69, 127, 28 and 23 eligible source cells, respectively. These descriptive scores evaluate alignment with annotated-stage expression displacements, rather than measured RNA derivatives. No direction reversal or UMAP smoothing enters the scores. Display streamlines use only the archived projected vectors, interpolated on a $100\times84$ grid with 35 neighbors, a Gaussian bandwidth of 0.42 UMAP units and a nearest-cell support radius of 0.40.

The marker analysis uses observed spliced counts for \textit{Neurog3}, \textit{Fev}, \textit{Chga}, \textit{Gcg}, \textit{Ins2}, \textit{Sst} and \textit{Ghrl}. Counts are normalized to 10,000 spliced transcripts per cell using the full raw gene matrix and transformed with $\log(1+x)$. The heatmap shows means in 24 equal-width bins of seed-0 pseudotime, scaled to $[0,1]$ separately for each gene; bins contain at least 41 cells. Binned peak locations are descriptive expression summaries. The seven genes are absent from the decoder panel and present in the original 2,000-gene graph panel. The resulting evidence concerns within-dataset stage and marker concordance; the original cell-state annotations provide a reference distinct from the earlier TSvelo ordering.
\subsection{Component ablations, timing and statistical summaries}
The displayed ablations change the U-dynamics residual, neighborhood information or joint fitting while fixing architecture settings, roots, splits and seeds. Removing U dynamics deletes only the U-ODE residual, retaining U reconstruction and the S equation with the original loss normalization. Removing neighborhood information replaces both neighbor-message operations with self features and sets the graph geometry penalty to zero. Separate fitting uses 2,700 expression updates followed by 2,400 kinetic updates, matching each module's update count in the full model while giving 5,100 total sequential updates. U2OS scores pool the three S-only folds within each seed; WT2 uses its all-cell expression fit.

Means and sample standard deviations summarize optimization seeds 0, 1 and 2. Paired changes compare matched seeds; these quantify optimization variability rather than biological replication. Simulations and full-S/U comparator rows use their recorded seed-0 results. Full recovery time includes initialization, target-data fitting and native outputs from prepared input matrices, excluding raw-read processing, software installation and figure rendering. Dagger-marked times sum measured, non-overlapping stages. CPU worker/thread counts and GPU use are reported alongside each time. Model-only timing repeats retain the original accuracy predictions. Hardware and parallelism differences are therefore visible in the comparison rather than absorbed into a common speed ranking.

\section*{Acknowledgements}
This work was supported by the National Natural Science Foundation of China (grant nos.\ 62573293 and 62473257) and the Science and Technology Commission of Shanghai Municipality (grant nos.\ 24ZR1435300 and 24510714300).

\section*{Data availability}
The public datasets analysed in this study are available from their original repositories. The WT1 and WT2 neurosphere data are deposited in the Gene Expression Omnibus under accession \href{https://www.ncbi.nlm.nih.gov/geo/query/acc.cgi?acc=GSE171636}{GSE171636}, and the U2OS FUCCI data under \href{https://www.ncbi.nlm.nih.gov/geo/query/acc.cgi?acc=GSE146773}{GSE146773}. The RPE1 scEU-seq data used for the supplementary preset inventory are part of \href{https://www.ncbi.nlm.nih.gov/geo/query/acc.cgi?acc=GSE128365}{GSE128365}. Dentate gyrus data are available from the \href{https://pklab.med.harvard.edu/velocyto/DentateGyrus/}{velocyto data repository}. The pancreatic endocrinogenesis and gastrulation erythroid datasets are distributed through the \href{https://scvelo.readthedocs.io/en/stable/scvelo.datasets.pancreas.html}{pancreas} and \href{https://scvelo.readthedocs.io/en/stable/scvelo.datasets.gastrulation_erythroid.html}{gastrulation\_erythroid} loaders in scVelo. The processed LARRY dataset is available on figshare at \href{https://doi.org/10.6084/m9.figshare.20780344}{doi:10.6084/m9.figshare.20780344}. Dataset subsets and evaluation references are specified in Supplementary Table S1. Simulation definitions and parameter settings are described in Methods and the Supplementary Information. Numerical comparison and ablation results are supplied with this manuscript as Supplementary Data~1, together with a dataset access list and file checksums.

\section*{Code availability}
The ManifoldVelo implementation, installation instructions, model configurations, runnable examples and software tests are available at \url{https://github.com/ZHOUZHEN2002/ManifoldVelo}. A versioned software release is provided as \href{https://github.com/ZHOUZHEN2002/ManifoldVelo/releases/tag/v1.0.0}{v1.0.0} (commit \texttt{581aa1f}). The package supports paired unspliced/spliced RNA fitting and pseudotime and velocity prediction for additional cells measured with spliced RNA alone. Original ManifoldVelo contributions are distributed under the Academic Research Only licence for non-commercial academic research, teaching and evaluation. The original TSvelo MIT notice and other third-party licence terms are retained in the repository.

\par
\end{document}